\documentstyle[prb,aps,twocolumn]{revtex}
\begin{document}
\input{epsf.tex}
\draft
%\preprint{}
\wideabs{
\title{Modeling of the Magnetic Susceptibilities of the
\protect\\ Ambient- and High-Pressure Phases of
(VO)$_{\bbox{2}}$P$_{\bbox{2}}$O$_{\bbox{7}}$}
\author{D.~C. Johnston}
\address{Ames Laboratory and Department of Physics and Astronomy, Iowa
State University, Ames, Iowa 50011}
\author{T. Saito, M. Azuma,
and M. Takano}
\address{Institute for Chemical Research, Kyoto University, Uji,
Kyoto-fu 611-0011, Japan}
\author{T. Yamauchi and Y. Ueda}
\address{Institute for Solid State Physics, University of Tokyo,
7-22-1 Roppongi, Minato-ku, Tokyo 106-8666, Japan}
\date{\today}

\maketitle

\begin{abstract}\hglue0.15in The magnetic susceptibilities $\chi$ versus
temperature $T$ of powders and single crystals of the ambient-pressure
(AP) and high-pressure (HP) phases of (VO)$_2$P$_2$O$_7$ are analyzed
using an accurate theoretical prediction of $\chi(T,J_1,J_2)$ for the
\mbox{spin-1/2} antiferromagnetic alternating-exchange ($J_1$,
$J_2$) Heisenberg chain.  The results are consistent with recent
models with two distinct types of alternating-exchange
chains in the AP phase and a single type in the HP phase.  The spin gap
for each type of chain is derived from the  respective set of two fitted
alternating exchange constants and the one-magnon dispersion relation for
each of the two types of chains in the AP phase is predicted.  The
influences of interchain coupling on the derived intrachain exchange
constants, spin gaps, and dispersion relations are estimated using a
mean-field approximation for the interchain coupling.  The accuracies of
the spin gaps obtained using fits to the low-$T$ $\chi(T)$ data by
theoretical low-$T$ approximations are determined.  The results of these
studies are compared with previously reported estimates of the exchange
couplings and spin gaps in the AP and HP phases and with the magnon
dispersion relations in the AP phase measured previously using inelastic
neutron scattering.
\end{abstract}
%\vspace{1cm}
}

\section{Introduction}
\vglue0.07in
A resurgence of research on the magnetic properties of low-dimensional
quantum spin systems has occurred over the last decade.  This work was
mainly initially undertaken to understand the relationships between
the magnetic properties of layered cuprates containing Cu$^{+2}$ spin-1/2
square lattice layers and the high superconducting transition
temperatures of the doped materials.\cite{Johnston1997}  This goal has
also spawned much research on related one- and two-dimensional (1D and
2D) spin systems.  Indeed, the new subfield of spin ladder physics was
created as a result of these efforts.\cite{LadderReviews,Johnston2000} 
The basic $n$-leg spin ladder consists of a planar arrangement of $n$
parallel vertical spin chains (the ladder ``legs'') with 
(nonfrustrating) horizontal nearest neighbor couplings between
adjacent chains, i.e., across the ladder ``rungs''.  

Self-doped two-leg spin ladders are realized in the compound ${\rm
NaV_2O_5}$, in which the V atoms are crystallographically equivalent and
the oxidation state of the V atoms is +4.5, resulting formally in a
mixed-valent $d^{0.5}$ system.\cite{NaV2O5}  However, the material is a
semiconductor rather than a metal.  Theoretical studies have indicated
that the reason for this is that the one $d$ electron per two V atoms is
localized on the respective V-O-V rung of the ladder due to the on-site
Coulomb repulsion on each rung.  This in turn led to the hypothesis that
each rung acts as a spin-1/2 site, in which case the compound is expected
to 
% Figure 1
\begin{figure}
\epsfxsize=2.8in
\centerline{\epsfbox{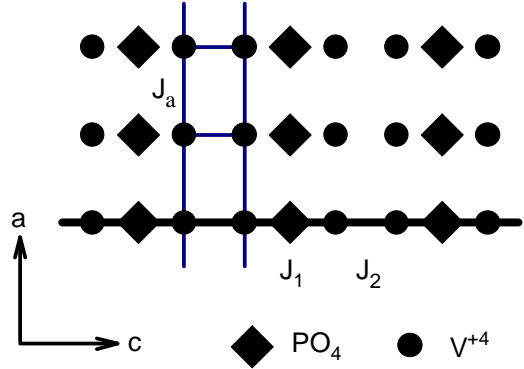}}

\caption{Sketch of the basic structures of AP-(VO)$_2$P$_2$O$_7$ and
HP-(VO)$_2$P$_2$O$_7$ in the $a$-$c$ plane, showing the exchange constants
$J_1$ and $J_2$ along the spin-1/2 V$^{+4}$ alternating-exchange chains
and $J_a$ in a perpendicular direction along the legs of the structural
two-leg ladders (adapted from Ref.~\protect\onlinecite{Garrett1997a}). 
The lattice parameters of AP-(VO)$_2$P$_2$O$_7$ are $a \approx 7.74$, $b
\approx 9.59$, and $c \approx 16.59$\,\AA, and of HP-(VO)$_2$P$_2$O$_7$
are $a \approx 7.58$, $b \approx 9.55$, and $c \approx 8.36$\,\AA.}
\label{VOPOStruct2}
\end{figure}
\vglue0.0in
\noindent behave magnetically like a spin $S = 1/2$ antiferromagnetic
(AF) Heisenberg chain.  Additionally, the observation of a spin
dimerization transition below 34\,K, accompanied by a lattice distortion,
is consistent with this scenario.  These and other aspects of the thermal
and magnetic behaviors of ${\rm NaV_2O_5}$ were recently examined in
detail in a combined theoretical and experimental
study.\cite{Johnston1999}  As part of this study, an accurate function
was generated for the theoretical magnetic susceptibility $\chi$ versus
temperature $T$ of the spin $S = 1/2$ AF alternating-exchange ($J_1$ and
$J_2$ with $J_2 \leq J_1$) Heisenberg chain over the entire range $0 \leq
\alpha
\leq 1$ of the alternation parameter
$\alpha\equiv J_2/J_1$.  

The availability of this high-accuracy theoretical $\chi(k_{\rm
B}T/J_1,\alpha)$ function now allows one to accurately and precisely test
the consistency of proposals for the occurrence of alternating-exchange
chains in specific compounds by comparing the observed $\chi(T)$ with
that expected theoretically.  In particular, in the present work we use
\ this \ method \ to \ test \ the \ consistency \ of \ recently 
\widetext
% Table I
\begin{table}
\caption{Spectroscopic splitting factor ($g$-factor) $g$, exchange
constants $J_1$ and $J_2$ and spin gap $\Delta$ determined by the
method listed for the ambient-pressure form of ${\rm (VO)_2P_2O_7}$ 
[AP-${\rm (VO)_2P_2O_7}$], for the high pressure form HP-${\rm
(VO)_2P_2O_7}$ and for the V dimer compound ${\rm (VO)HPO_4\cdot{1\over
2}H_2O}$.  Quantities marked with a daggar ($^\dag$) are derived using
the $S = 1/2$ alternating-exchange chain model whereas those marked by an
asterisk~(*) are derived using the $S = 1/2$ two-leg ladder model. 
Quantities with no mark make no assumption about the specific model
and/or are determined or assumed independently of the fits used to obtain
the other listed quantities.  All measurements were carried out on
polycrystalline samples except as otherwise noted.  The listed alternating
exchange constants from Ref.~\protect\onlinecite{Uhrig1998} were obtained
within a 2D coupled alternating-exchange chain model with interchain
couplings given in the text and were determined from a fit to
one-magnon inelastic neutron scattering data
(Ref.~\protect\onlinecite{Garrett1997a}) with the spin gap listed and from
high-$T$ $\chi(T)$ data (Ref.~\protect\onlinecite{Johnston1987}).  The
error bars on the parameters listed from the present work include the
estimated influences of interchain interactions.}
\begin{tabular}{lccccccr}
Compound & $g$ & $\alpha$ & $J_1/k_{\rm B}$\,(K) & $J_2/k_{\rm B}$\,(K) &
$\Delta/k_{\rm B}$\,(K) & Method & Ref.\\
\hline
AP-${\rm (VO)_2P_2O_7}$ & 2.00$^\dag$ & 0.7$^\dag$ & 131$^\dag$ &
92$^\dag$ & & $\chi(T)$  &\onlinecite{Johnston1987} \\
                       & 1.99$^\dag$ & 0.722$^\dag$ & 128.9$^\dag$ &
93.0$^\dag$ & 57$^\dag$ & $\chi(T)$  &\onlinecite{Barnes1994} \\
                       & 2.03$^\ast$ & 1.007$^\ast$ & 90.6$^\ast$ &
90.0$^\ast$ & 45.7$^\ast$ & $\chi(T)$  &\onlinecite{Barnes1994} \\
&&&&& 30    & $^{31}K(T)$ & \onlinecite{Furukawa1996} \\
&&&&& 60    & $^{31}(1/T_1)(T)$ & \onlinecite{Furukawa1996} \\
&&&&& 34(3)    & $^{31}K(T)$ & \onlinecite{Kikuchi1997} \\
&&&&& 100    & $^{31}(1/T_1)(T)$ & \onlinecite{Kikuchi1997} \\
&&&&& 52    & $\chi(T)$ & \onlinecite{Schwenk1996} \\
&&&&& 43(2) & $^0n$ scatt. & \onlinecite{Eccleston1994} \\
~~~crystal& 1.937&&&& 35 & $c_{11}(H,1.6$\,K) & \onlinecite{Wolf2000}
\\ ~~~crystal&&&&& 36 & Raman (low $T$) & \onlinecite{Grove1999} \\
&&&&& 40.4(4), 70 & $^0n$ scatt. & \onlinecite{Garrett1997b} \\
~~~crystals&&&&& 36.2(3), 66.7(2) & $^0n$ scatt. &
\onlinecite{Garrett1997a}
\\
 & 1.97 &  &  &  & 33(1), 62(3) & $M(H$,1.3\,K) &
\onlinecite{Yamauchi1999}
\\
 & &  &  &  & 35(2), 52(3) & $^{31}K(T)$ & \onlinecite{Yamauchi1999} \\
 &&&&& 53, 71 & $^{31}(1/T_1)(T)$ & \onlinecite{Kikuchi1999} \\
 &&&&& 68(2) & $^{51}K(T)$ & \onlinecite{Kikuchi1999} \\
 &&&&& 66(2) & $^{51}(1/T_1)(T)$ & \onlinecite{Kikuchi1999} \\
 &&0.83$^\dag$, & 124$^\dag$, & 103$^\dag$, & 35, & Inferred  &
\onlinecite{Kikuchi1999} \\
 &&0.67$^\dag$ & 136$^\dag$ & 92$^\dag$ & 68 \\
&& 0.793$^\dag$ & 124$^\dag$ & 99$^\dag$ &36.2(3)& Theory &
\onlinecite{Uhrig1998} \\ 
~~~crystal& 2 & 0.67$^\dag$ & 122$^\dag$ & 82$^\dag$ & 58$^\dag$ &
$\chi(T\approx 70$\,K) &
\onlinecite{Prokofiev1998} \\
~~~crystal&&&&& 67 & ESR $I(T$) & \onlinecite{Prokofiev1998} \\
& 2.021(2)$^\dag$ & 0.782(2)$^\dag$ &
130.9(5)$^\dag$ & 102.4(6)$^\dag$ &  & $\chi(T)$ & \onlinecite{Hiroi1999}
\\
&  & 0.85(1)$^\dag$, & 135(5)$^\dag$, &
115(6)$^\dag$, & 38.4(9)$^\dag$, & $\chi(T)$ & This work \\
&  & 0.638(7)$^\dag$ & 127(3)$^\dag$ & 81(3)$^\dag$ & 67(1)$^\dag$ \\
\hline
HP-${\rm (VO)_2P_2O_7}$ & 2 &&&& 23 & $M(H,$1.3\,K) &
\onlinecite{Azuma1999}
\\  & & & & & 27 & $\chi(T< 30$\,K) & \onlinecite{Azuma1999} \\ 
& 2.01$^\dag$ & 0.9$^\dag$ & 137$^\dag$ & 123$^\dag$ & 27$^\dag$ &
$\chi(T > 30$\,K) &
\onlinecite{Azuma1999} \\ 
~~~crushed crystals&& 0.8737(13)$^\dag$ & 135.6(7)$^\dag$ &
118.5(8)$^\dag$ & 33.9(2)$^\dag$ & $\chi(T)$ & This work \\ 
\hline
${\rm (VO)HPO_4\cdot{1\over 2}H_2O}$ & 1.99 & 0 & 88.0 & 0 & 88.0 &
$\chi(T)$ & \onlinecite{Johnson1984} \\
&&&&& 74 & $^{31}K(T)$ & \onlinecite{Furukawa1996} \\
& & 0 & 90.6(5) & 0 & 90.6(5) & $^0n$ scatt. & \onlinecite{Tennant1997}
\end{tabular}
\label{TabJAPVOPO}
\end{table}
\noindent 
\narrowtext
\noindent proposed alternating-exhange chain models for the ambient- and
high-pressure phases of vanadyl pyrophosphate, ${\rm (VO)_2P_2O_7}$, and
we obtain the exchange constants and spin gaps in the respective chains. 
The influence of interchain coupling on the derived intrachain exchange
constants and spin gaps is investigated using a mean-field approximation
for the interchain coupling.  The accuracies of the spin gaps obtained
using fits to the low-$T$ $\chi(T)$ data by theoretical low-$T$
approximations are determined.  The results of these studies are compared
with previously reported estimates of the exchange couplings and spin
gaps in the two phases and with the dispersion \  relations \
measured \ for \ the \ ambient-pressure 

\newpage
\noindent 
phase by inelastic neutron scattering.

The history of the study of the magnetic properties of ${\rm
(VO)_2P_2O_7}$ is interesting and extensive.  In the following two
sections we give brief overviews of the previous work on the
ambient-pressure (AP) and high-pressure (HP) phases of this compound,
respectively, and then present the plan for the remainder of the paper.

\subsection{AP-$\bbox{\rm (VO)_2P_2O_7}$}

\vglue0.13in

The V$^{+4}\ d^1$ ambient-pressure phase AP-${\rm (VO)_2P_2O_7}$,
sometimes abbreviated in the recent literature as ``VOPO'', is an
industrial catalyst for the selective oxidation of $n$-butane to maleic
anhydride.\cite{Johnson1987,Gai1995}  This compound was found to have an
orthorhombic crystal structure\cite{Gorbunova1979,Johnson1984} (containing
four inequivalent types of V atoms) which can be viewed
crystallographically as containing $S = 1/2$ two-leg
ladders,\cite{Johnston1987} where the rungs of the ladder lie along the
$c$-axis and the legs are oriented along the $a$-axis, as sketched in
Fig.~\ref{VOPOStruct2}.\cite{Garrett1997a}  A single crystal x-ray
diffraction structural study claimed that the previous structural studies
were incorrect and that the structure is monoclinic with eight
inequivalent V atoms in the unit cell, although the overall structural
features and the unit cell dimensions were found to be very similar to
those of the previously proposed orthorhombic
structure.\cite{Nguyen1995}   However, a recent study of a
polycrystalline sample using both x-ray and neutron diffraction Rietveld 
refinements and transmission electron microscopy confirmed the
orthorhombic structure and ruled out the monoclinic
structure;\cite{Hiroi1999} it was suggested that whether the orthorhombic
or monoclinic structure occurs in a particular sample may depend on the
exact composition and the details of sample synthesis.\cite{Hiroi1999} 
There have been two conventions used in the literature for designating the
$b$ and $c$ axes in which these two axes are mutually interchanged; we
will adhere to the convention in Fig.~\ref{VOPOStruct2}, as in
Ref.~\onlinecite{Hiroi1999}, for which the approximate lattice parameters
are listed in the figure caption.

\begin{figure}
\epsfxsize=3.4in
\centerline{\epsfbox{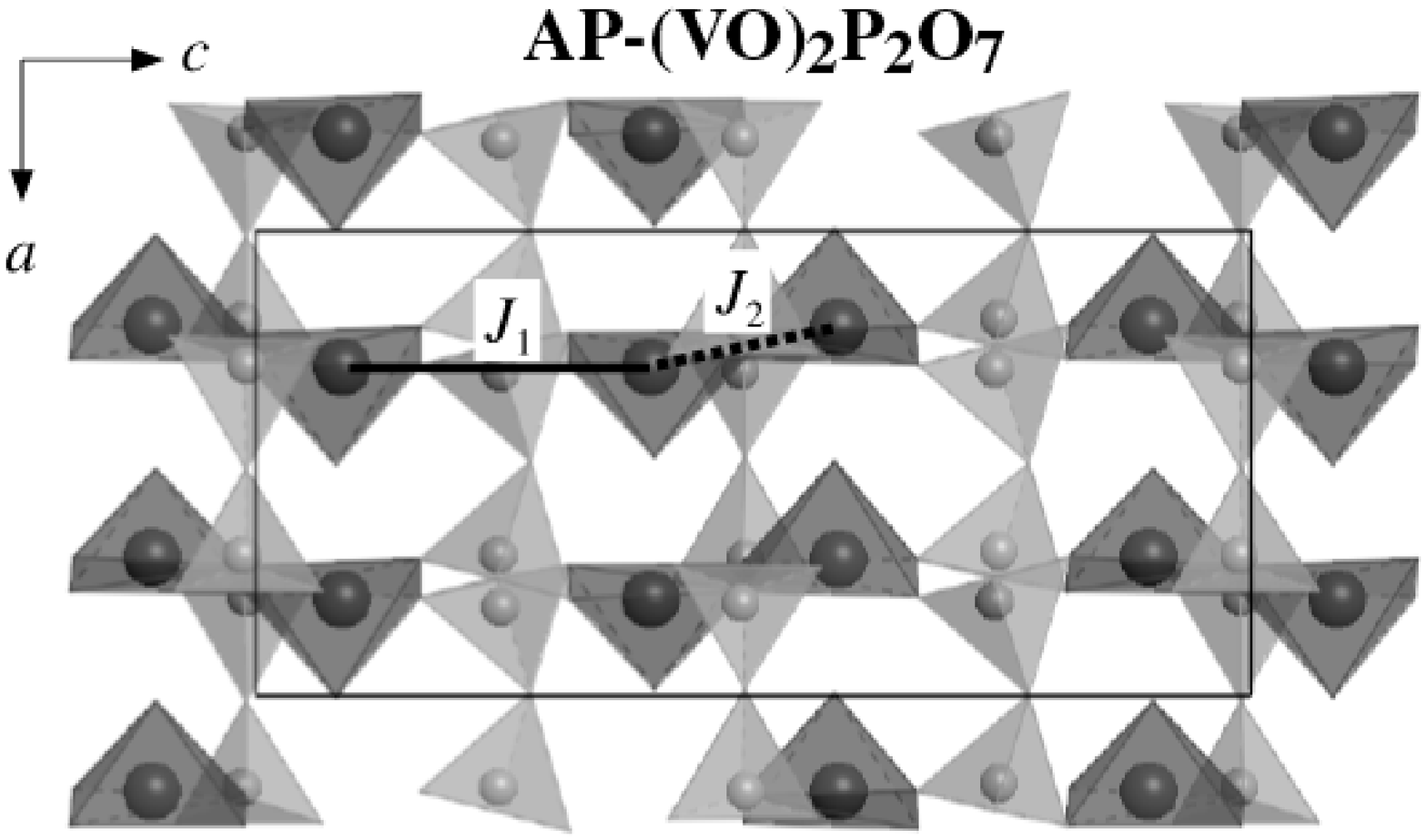}}
\vglue 0.1in
\epsfxsize=3.2in
\centerline{\epsfbox{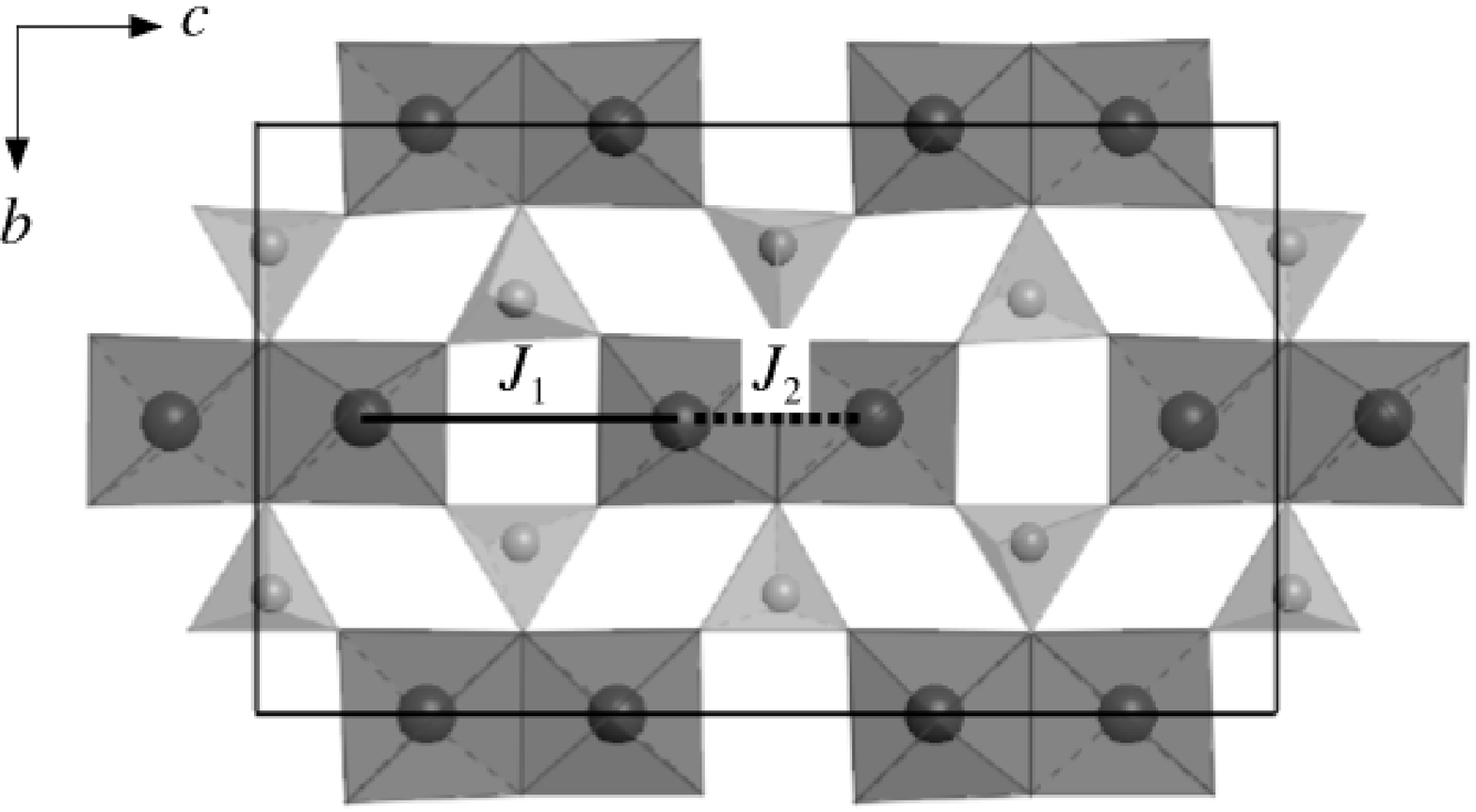}}
\vglue 0.2in
\caption{Detailed crystal structure of AP-${\rm (VO)_2P_2O_7}$ as viewed
along the $b$-axis (top panel) and along the $a$ axis (bottom panel).  The
rectangular boxes are outlines of the unit cell in the
respective planes.  The large spheres are V atoms and the small spheres
are P atoms.  Oxygen atoms (not shown) are at the vertices of the VO$_5$
square pyramids and the PO$_4$ tetrahedra.  The pairs of V atoms coupled
by exchange constants $J_1$ and $J_2$ along the $S = 1/2$ AF
alternating-exchange Heisenberg chains are shown.}
\label{AP-VOPOStruct}
\end{figure}

The magnetic susceptibility versus temperature $\chi(T)$ of AP-${\rm
(VO)_2P_2O_7}$ was found to exhibit an energy gap (``spin gap'') for
magnetic excitations.\cite{Johnston1987}  The $\chi(T)$ was initially
fitted to high precision by the prediction for the $S = 1/2$
AF alternating-exchange Heisenberg chain, with the
exchange constants $J_1$ and $J_2$ and alternation parameter $\alpha
\equiv J_2/J_1$  given in Table~\ref{TabJAPVOPO}.\cite{Johnston1987}  A
fit by the spin ladder model as suggested\cite{Johnston1987} from the
crystal structure (see Figs.~\ref{VOPOStruct2} and~\ref{AP-VOPOStruct})
was not possible at that time (1987) due to lack of theoretical
predictions for
$\chi(T)$ of this model.  It was also speculated that even though the
crystallographic features suggest a spin ladder model, the actual
magnetic interactions might turn out to correspond to those of
alternating-exchange spin chains.\cite{Johnston1987}  When  $\chi(T)$
calculations for the spin ladder model were eventually
done,\cite{Barnes1994} it was found that the same experimental $\chi(T)$
data set\cite{Johnston1987} could be fitted by the spin ladder model to
the same high precision as for the very different alternating-exchange
chain model.\cite{Barnes1994}  The existence of a spin gap was
subsequently confirmed and its value estimated from NMR measurements of
the $^{31}$P Knight shift $^{31}K(T)$ and nuclear spin-lattice relaxation
rate $^{31}(1/T_1)(T)$,\cite{Furukawa1996,Kikuchi1997} from $\chi(T)$
(Ref.~\onlinecite{Schwenk1996}) and inelastic neutron
scattering\cite{Eccleston1994} measurements on polycrystalline samples at
low temperatures, and most recently from elastic constant
measurements\cite{Wolf2000} on a single crystal in pulsed magnetic fields
up to 50\,T at $T = 1.6$\,K and from Raman scattering intensity
measurements on single crystals at low temperatures,\cite{Grove1999} as
listed in Table~\ref{TabJAPVOPO}.  

The above inelastic neutron scattering measurements on polycrystalline
AP-${\rm (VO)_2P_2O_7}$ reportedly confirmed the spin ladder model and
rejected the alternating-exchange chain model by a comparison of the
observed spin gap [43(2)\,K] with the values 45.7\,K and
57\,K predicted for the two respective models from the sets of exchange
constants determined from respective fits to the $\chi(T)$
data.\cite{Eccleston1994}  However, subsequent inelastic neutron
scattering results on polycrystalline samples\cite{Garrett1997b} and
especially on a collection of about 200 oriented small single
crystals\cite{Garrett1997a} proved that AP-${\rm (VO)_2P_2O_7}$ is not a
spin-ladder compound.  The strongest dispersion of the one-magnon spectra
of the single crystals was found to be along the $c$ axis, i.e., in the
direction of the structural ladder rungs, and the coupling in the
direction of the ladder legs was found to be weakly
ferromagnetic.\cite{Garrett1997a}  Thus, perhaps surprisingly, the
superexchange coupling path \mbox{V-O-P-O-V} along the
$c$ axis, coupling the structural two-leg ladders as shown in
Fig.~\ref{AP-VOPOStruct}, is much stronger than the shorter \mbox{V-O-V}
coupling along the ladder legs parallel to the $a$-axis.  These results
were interpreted in terms of an alternating-exchange chain model with the
chains running along the $c$ axis, with weak coupling between the chains,
thus confirming the above speculation in Ref.~\onlinecite{Johnston1987}. 
Subsequent $\chi(T)$ data for powder\cite{Hiroi1999} and single
crystal\cite{Prokofiev1998} samples have been interpreted in terms of the
alternating-exchange chain model, with exchange parameters listed in
Table~\ref{TabJAPVOPO}.

The spin gaps of 43(2)\,K (Ref.~\onlinecite{Eccleston1994}) and
40.4(4)\,K (Ref.~\onlinecite{Garrett1997b}) found from inelastic
neutron scattering measurements on powder samples are both significantly
larger than the one-magnon spin gap of 36.2(3)\,K determined from the
neutron scattering measurements on single crystals.\cite{Garrett1997a} 
Since powder samples have usually been found to show relatively high
levels of paramagnetic impurities and/or defects, this comparison suggests
that the larger spin gaps in the powder samples may be a real effect
arising from termination of the spin chains by defects.  One
would indeed expect finite segments of alternating-exchange chains to
exhibit larger spin gaps than for the infinite chain.

We have also listed in \ Table~\ref{TabJAPVOPO} \ the \ intradimer \
exchange
\ constant \ determined \ from \ $\chi(T)$ \
(Ref.~\onlinecite{Johnson1984}),
\ $^{31}$P NMR \ Knight \ shift \ $^{31}K(T)$ 
\ (Ref.~\onlinecite{Furukawa1996}), \ and \ inelastic \ neutron
\ scattering\cite{Tennant1997} \ measurements
\ of \ polycrystalline \ samples \ of \ the \ $d^1$ \ $S = 1/2$ \ vanadium
\ dimer \ compound
% Figure 2
\vglue0.4in
\noindent
${\rm (VO)HPO_4\cdot{1\over 2}H_2O}$ (vanadyl hydrogen phosphate
hemihydrate), which is a precursor for the synthesis of, and has
structural similarities to, AP-${\rm
(VO)_2P_2O_7}$.\cite{Johnson1987,Johnson1984}  In particular, the neutron
scattering study of this compound confirmed the importance and strength
of the \mbox{V-O-P-O-V} superexchange pathway.\cite{Tennant1997}

From the above neutron scattering measurements on single crystals of
AP-${\rm (VO)_2P_2O_7}$, a second spin gap at a larger energy of 67\,K
was found in addition to the gap of 36\,K for coherent one-magnon
propagation along the $c$ axis.\cite{Garrett1997a}  The two spin gaps
cannot both arise from one-magnon excitations in an isolated
alternating-exchange chain and and the larger one was  suggested to arise
from neutron scattering from two-magnon triplet bound states of such
chains, although the scattered neutron intensity was larger than expected
from scattering from such states.\cite{Garrett1997a}  A 2D model
incorporating both  nonfrustrating ($J_a$, see Fig.~\ref{VOPOStruct2})
and diagonal frustrating ($J_\times$) AF interactions between
alternating-exchange chains was subsequently proposed.\cite{Uhrig1998} 
\ The \ intrachain \ and interchain ex-
% Figure 3
\begin{figure}
\epsfxsize=2.4in
\centerline{\epsfbox{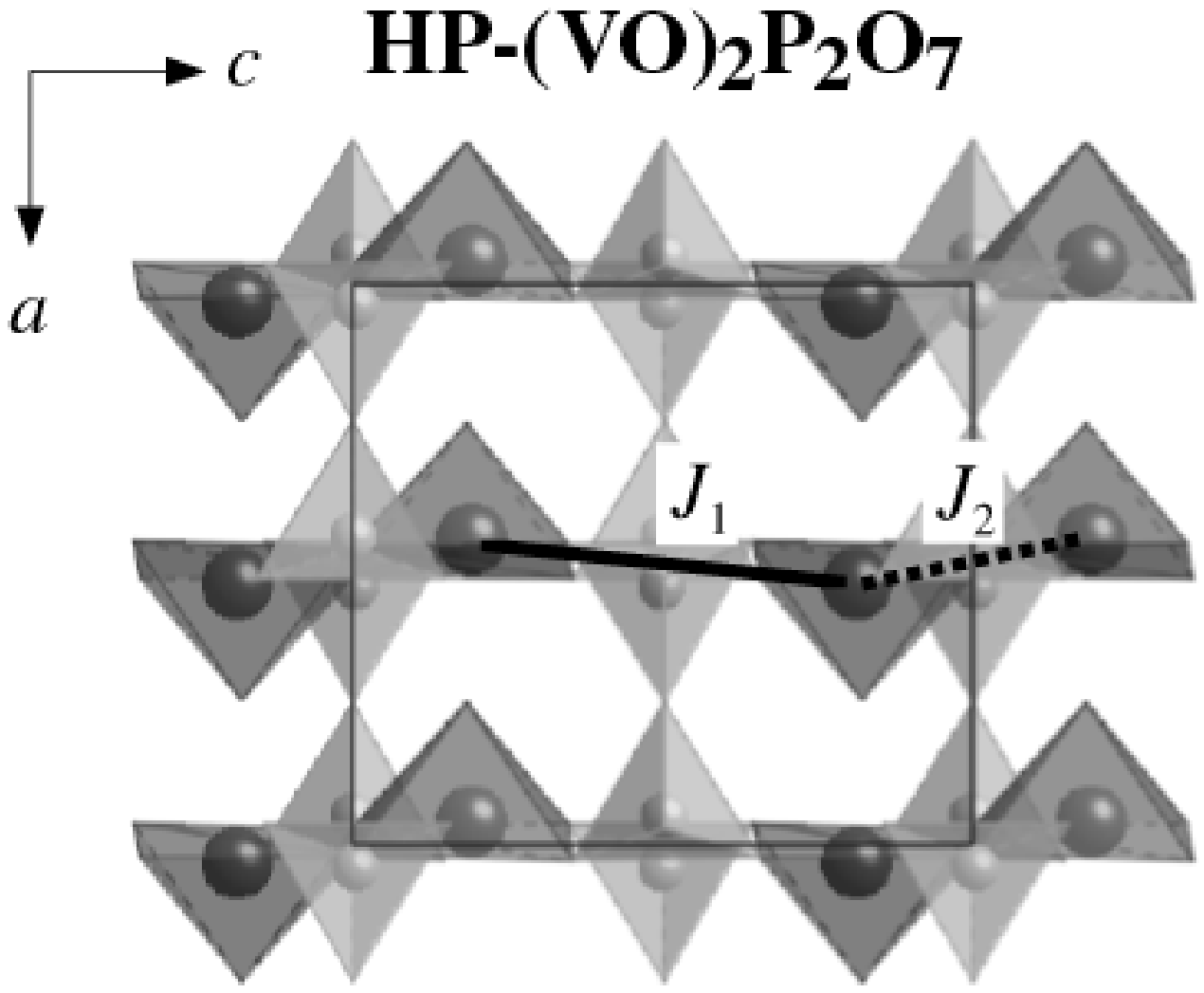}}
\vglue 0.2in
\epsfxsize=2.4in
\centerline{\epsfbox{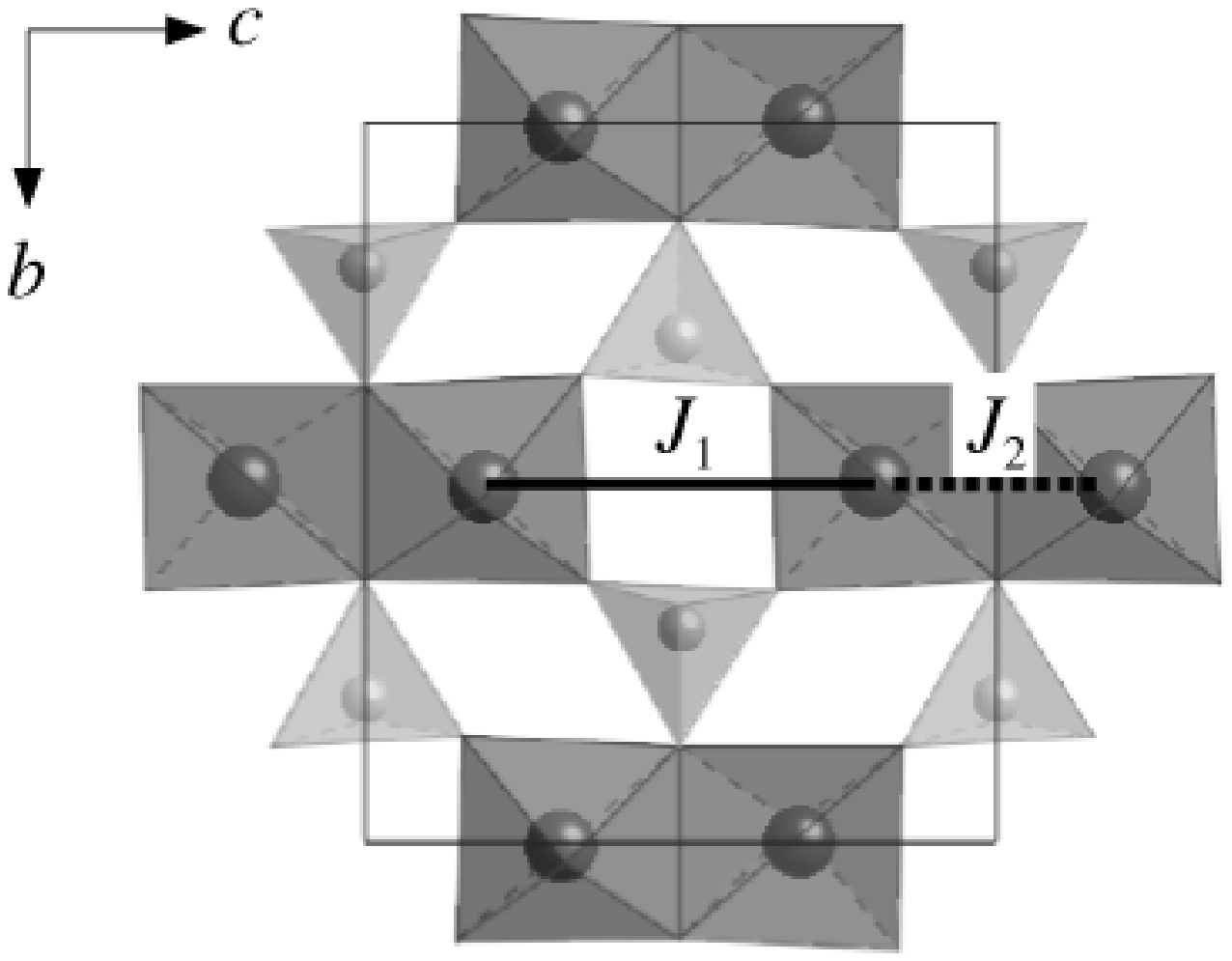}}
\vglue 0.1in
\caption{Detailed crystal structure of HP-${\rm (VO)_2P_2O_7}$.  The
designations are the same as in Fig.~\protect\ref{AP-VOPOStruct}.}
\label{HP-VOPOStruct}
\end{figure}
\vglue0.66in
\noindent change constants were determined by a
(very good) fit to the one-magnon dispersion relation (including the
one-magnon spin gap of 36\,K) measured in the inelastic
neutron scattering measurements and by using the
results\cite{Johnston1987} of high-$T$ $\chi(T)$ measurements.  The two
alternating AF intrachain exchange constants so
determined\cite{Uhrig1998} are listed in Table~\ref{TabJAPVOPO}.  The two
AF interchain interactions were predicted to be rather large: $J_a/J_1 =
0.203$ and $J_\times/J_1 = 0.255$.\cite{Uhrig1998}  However, the
$\chi(T)$ calculated using this model in Ref.~\onlinecite{Uhrig1998} was
found to be in poor agreement with experimental single-crystal $\chi(T)$
data\cite{Prokofiev1998} even at the highest measured temperature $T\sim
200$\,K, which is significantly above the temperature ($\sim 70$\,K)
at which $\chi(T)$ shows a broad maximum and where the prediction would be
expected to be quite accurate.  On the other hand, we find in
Sec.~\ref{SecAP-VOPOMFT} below that their $\chi(T)$ prediction is in
quite good agreement with experimental data if the comparison is done in
a somewhat different way.  Additional calculations indicated that the
frustrating AF interchain interaction stabilizes two-magnon bound states
and thus supported the conjecture\cite{Garrett1997a} that the higher
energy mode at 67\,K is a triplet two-magnon bound
state.\cite{Uhrig1998}  The same 2D model containing alternating-exchange
chains and frustrating AF interchain couplings and also a model
containing ferromagnetic interchain couplings were studied in
Ref.~\onlinecite{Weisse1999} where the former model was found to agree
better with the experimental neutron scattering data.  From a Raman
scattering study of single crystals, the spin-phonon interaction was
suggested to be responsible for the formation of the two-magnon triplet
bound states identified in the neutron scattering experiments, rather
than arising from frustrating interchain interactions.\cite{Grove1999}

Electron spin resonance (ESR) intensity versus temperature $I(T)$ data
obtained for a single crystal were interpreted in terms of an isolated AF
spin dimer model, yielding a spin gap of 67\,K,\cite{Prokofiev1998}
corresponding to the higher-energy gap seen in the inelastic neutron
scattering data.  In this ESR study, transitions between the Zeeman
levels within the one-magnon triplet band were reportedly not observed. 
These $I(T)$ data were subsequently  re-interpreted within the
above 2D coupled alternating-exchange chain model as arising from
transitions between the Zeeman levels within the one-magnon band and good
agreement between the theoretical prediction for $I(T)$ and the
experimental $I(T)$ data was found.\cite{Uhrig1998}  

On the other hand, recent $^{31}$P and $^{51}$V NMR and magnetization
versus applied magnetic field $M(H)$ measurements at high fields and
low temperatures have indicated that there are two magnetically distinct
types of alternating-exchange V chains in AP-${\rm (VO)_2P_2O_7}$,
interpenetrating with each other, each with its own spin
gap.\cite{Yamauchi1999,Kikuchi1999}  The two spin gaps inferred for the
two types of chains, 35(2)\,K from $^{31}K(T)$
measurements\cite{Yamauchi1999} and 68(2)\,K from $^{51}$V
$K(T)$ measurements,\cite{Kikuchi1999} and 33(1)\,K and 62(3)\,K from
$M(H)$ measurements at 1.3\,K,\cite{Yamauchi1999} agree well with the
above two spin gaps found from the neutron scattering measurements, 
respectively, thus providing an alternate explanation for the larger of
the two spin gaps.  Using additional information from the neutron
scattering one-magnon dispersion relation
measurements,\cite{Garrett1997a} the two alternating exchange constants
in each chain have been estimated\cite{Kikuchi1999} and are listed in
\mbox{Table}~\ref{TabJAPVOPO}.  This model is not supported by the Raman
scattering results.\cite{Grove1999}  However, recent unpublished 
inelastic neutron scattering measurements on a large single crystal at $T
= 2$\,K in zero and high magnetic fields show that the two types of
magnetic excitations found earlier\cite{Garrett1997a} split in a magnetic
field according to expectation for two triplet bands, results which
constitute independent evidence for the validity of the two-chain model
for AP-${\rm (VO)_2P_2O_7}$.\cite{NaglerPC}  In addition, two
previously undetected magnetic excitations termed ``shadow bands''
were found which are thought to arise from the staggered alignment of
successive V dimers along the V alternating-exchange chains within the
structure (see the top panel of
Fig.~\ref{AP-VOPOStruct}).\cite{NaglerPC,Damle1999}

At present there is thus no universal agreement about a Hamiltonian which
can self-consistently explain the various experimental measurements
probing the magnetism in \mbox{AP-${\rm (VO)_2P_2O_7}$}, although all of
the models considered recently contain $S = 1/2$ AF alternating-exchange
Heisenberg chains as an \,essential \,element and a \,consen- 
\newpage 
\noindent sus is emerging
that the two-chain model can explain many of the observed properties.

\subsection{HP-$\bbox{\rm (VO)_2P_2O_7}$}

The high-pressure phase HP-${\rm (VO)_2P_2O_7}$ was recently synthesized
by heating polycrystalline \mbox{AP-${\rm (VO)_2P_2O_7}$} for 1/2\,h at
700\,$^\circ$C under a pressure of 2\,GPa.\cite{Azuma1999}  As shown in
Fig.~\ref{HP-VOPOStruct} and by comparison with Fig.~\ref{AP-VOPOStruct},
HP-${\rm (VO)_2P_2O_7}$ has a simpler structure than \mbox{AP-${\rm
(VO)_2P_2O_7}$}.  In \mbox{HP-${\rm (VO)_2P_2O_7}$}, all V atoms are
crystallographically equivalent but the same basic structure as in the
ambient-pressure phase was found.\cite{Hiroi1999,Azuma1999}  The
similarities between the two structures suggest that \mbox{HP-${\rm
(VO)_2P_2O_7}$} also contains $S = 1/2$ AF alternating-exchange chains,
but of a single type.\cite{Azuma1999}  Indeed, high-field $M(H)$
measurements at 1.3\,K revealed a single spin gap of $\sim 23$\,K,
consistent with this hypothesis.\cite{Azuma1999}  Modeling of $\chi(T)$
data below 30\,K was carried out using the low-$T$ approximation in
Eq.~(\ref{EqTroyer:a}) below for the spin susceptibility of a
1D spin system with a spin gap, yielding a similar spin gap of
27\,K\@.\cite{Azuma1999}  The $\chi(T)$ data above 30\,K were analyzed
using the spin susceptibility of the $S = 1/2$ alternating-exchange
Heisenberg chain model, yielding the exchange constants listed in
Table~\ref{TabJAPVOPO} and the same spin gap
$\Delta/k_{\rm B} = 27$\,K\@.\cite{Azuma1999}  These estimates of the gap
value are similar to the one-magnon gap of $\approx 36$\,K in
AP-${\rm (VO)_2P_2O_7}$ found from the inelastic neutron scattering and
other measurements discussed above.  

Equation~(\ref{EqTroyer:a}) has also been previously used to fit $\chi(T)$
data for other $S = 1/2$ 1D compounds, but to our knowledge all such
studies, with the exception of a study of the ${\rm Cu^{+2}}$ $S = 1/2$
two-leg Heisenberg spin ladders in ${\rm SrCu_2O_3}$,\cite{Johnston1996}
have assumed that $A$ and the spin gap $\Delta$ are independently
adjustable parameters when fitting experimental $\chi(T)$ data.  We
discuss in Sec.~\ref{SecThy} below that $A$ is uniquely related
to $\Delta$ and is not an independently adjustable
parameter for a given type of 1D spin lattice.\cite{Johnston1999}  On
the other hand, if one assumes that the spin lattice in a material has
a spin gap but the type of 1D spin lattice is unknown, or if the fit is
not carried out in the low-temperature limit, then $A$ and $\Delta$
would have to be treated as independently adjustable parameters.  In
the present work we evaluate the accuracy of using Eq.~(\ref{EqTroyer:a})
to determine the spin gap of the 1D $S = 1/2$ alternating-exchange
Heisenberg chain from $\chi(T)$ data for HP-${\rm (VO)_2P_2O_7}$ by
comparing the spin gap obtained using this approximation with the spin gap
obtained by modeling the same data set using the accurate theoretical 
prediction\cite{Johnston1999} for the spin susceptibility of the $S =
1/2$ AF alternating-exchange Heisenberg chain model.

Most recently, single crystals of HP-${\rm (VO)_2P_2O_7}$ have been
grown.\cite{Saito2000} The anisotropic spectroscopic splitting
factors ($g$ factors) of the $S = 1/2$ V$^{+4}$ ions in a single
crystal were determined from ESR measurements
and the anisotropic magnetic susceptibilites of a single crystal were 
measured.\cite{Saito2000}  Here we perform detailed modeling of these
$\chi(T)$ data using the $g$ factors determined from ESR and using the $S
= 1/2$ AF alternating-exchange Heisenberg chain model for the spin
susceptibility.  We  determine the exchange constants within the chains
and from these we obtain an estimate of the spin gap.  An independent
estimate of the spin gap is obtained by modeling only the low-temperature
data.

\subsection{Plan of the Paper}

The plan for the remainder of this paper is as follows.  In
Sec.~\ref{SecThy} a summary is given of the theory that we will need to
carry out and discuss the modeling described above.  General
considerations for fitting experimental data by theoretical predictions
for the spin susceptibility are discussed in Sec.~\ref{SecExpIntro}. 
New and literature $\chi(T)$ data for AP-${\rm (VO)_2P_2O_7}$ are
presented and fitted in Sec.~\ref{SecPowderChiAPVOPO}.  High-precision
fits to the $\chi(T)$ of a high-purity powder sample are presented in
Sec.~\ref{SecAPVOPOPowderChi}.  In this section, we show how the fitted
exchange constants and spin gap(s) of the model of alternating-exchange
chain(s) vary depending on whether a single-chain or two-chain model is
used to fit the data, and on whether the $g$ value is fixed or allowed to
vary during the fits.  The influences of interchain couplings on the
exchange constants and spin gaps inferred from modeling $\chi(T)$ for the
powder sample are quantitatively determined in Sec.~\ref{SecAP-VOPOMFT}
using a molecular field theory for the interchain couplings, where we also
compare our derived interchain couplings with the corresponding
theoretical predictions of Uhrig and Normand\cite{Uhrig1998} which were
obtained using a one-chain model.  The $\chi(T)$ data for two single
crystals of AP-${\rm (VO)_2P_2O_7}$ are analyzed using the two-chain
model in Sec.~\ref{SecAPVOPOXtalChi}.  In Sec.~\ref{SecAPVOPODispRelns}
we test our predicted dispersion relations for the two proposed chains by
comparison with the results of inelastic neutron scattering
measurements.  The $\chi(T)$ data for HP-${\rm (VO)_2P_2O_7}$ are
presented and modeled in Sec.~\ref{SecChiHPVOPO}.  The most accurate and
precise $\chi(T)$ data for this phase were obtained for a sample of
crushed crystals.  These data are analyzed using the one-chain model in
Sec.~\ref{SecCrushedXtalChi}.  The influences of interchain couplings on
the derived exchange constants and spin gap of this sample are considered
in Sec.~\ref{SecHPVOPOMFT}.  Our evaluation of the accuracy of the spin
gap obtained using a theoretical low-$T$ approximation to the spin
susceptibility of a 1D $S = 1/2$ spin system with a spin gap, as
previously used to analyze powder
$\chi(T)$ data for \mbox{HP-${\rm (VO)_2P_2O_7}$},\cite{Azuma1999} is
given in Sec.~\ref{SecPowderChi}.  The anisotropic $\chi(T)$
data\cite{Saito2000} for a single crystal of HP-${\rm (VO)_2P_2O_7}$ are
modeled in Sec.~\ref{SecXtalChi}\@.  The powder average of these single
crystal $\chi(T)$ data are modeled in Sec.~\ref{SecChiPwdAve} by two
low-$T$ approximations for the spin susceptibility to obtain an
independent estimate of the spin gap.  A summary of our modeling results
and our conclusions are given in Sec.~\ref{SecDisc}.

\section{Theory}
\label{SecThy}

The Hamiltonian for the $S = 1/2$ alternating-exchange Heisenberg chain is
written in three equivalent ways as
\begin{mathletters}
\label{EqAltChnHam:all}
\begin{eqnarray} {\cal H} & = & \sum_i J_1 \bbox{S}_{2i-1}\cdot
\bbox{S}_{2i} + J_2
\bbox{S}_{2i}\cdot \bbox{S}_{2i+1}\label{EqAltChnHam:a}\\ & = & \sum_i J_1
\bbox{S}_{2i-1}\cdot \bbox{S}_{2i} + \alpha J_1
\bbox{S}_{2i}\cdot \bbox{S}_{2i+1}\label{EqAltChnHam:b}\\ & = & \sum_i 
J(1 +
\delta) \bbox{S}_{2i-1}\cdot \bbox{S}_{2i} + J(1 -
\delta) \bbox{S}_{2i}\cdot \bbox{S}_{2i+1}~,\label{EqAltChnHam:c}
\end{eqnarray}
\end{mathletters} where
\begin{mathletters}
\label{EqDimParams:all}
\begin{eqnarray} J_1 & = &  J(1 + \delta) = \frac{2J}{1 +
\alpha}~,\label{EqDimParams:a}\\
\nonumber\\
\alpha & = & {J_2\over J_1} =
\frac{1-\delta}{1+\delta}~,\label{EqDimParams:b}\\
\nonumber\\
\delta & = & {J_1\over J} - 1 = {J_1 - J_2\over 2J} =
\frac{1-\alpha}{1+\alpha}~,\label{EqDimParams:c}\\ 
\nonumber\\ J & = & {J_1 + J_2\over 2} = J_1{1 + \alpha\over
2}~,\label{EqDimParams:d}
\end{eqnarray}
\end{mathletters} with AF couplings $J_1 \geq J_2 \geq 0$, $0 \leq
(\alpha,\,\delta) \leq 1$.  The uniform undimerized chain corresponds to
$\alpha = 1,\ \delta = 0$ and $J_1 = J_2 = J$, whereas the isolated dimer
has
$\alpha = 0,\ \delta = 1,\ J_2 = 0$ and intradimer exchange interaction
$J_1$.  The form of the Hamiltonian in Eq.~(\ref{EqAltChnHam:c}), in
which the appropriate variables are
$\delta$ and the average exchange constant along the chain $J$ instead of
$\alpha$ and the maximum exchange constant $J_1$ as in
Eq.~(\ref{EqAltChnHam:b}), is often used for compounds in which the spin
dimerization is weak and/or for systems showing a second-order spin
dimerization transition with decreasing temperature such as occurs in
spin-Peierls systems.

The spin gap $\Delta$ for magnetic spin excitations from the $S = 0$
ground state to the lowest-lying $S = 1$ triplet excited states for the
alternating-exchange chain is uniquely related to the alternation
parameter $\alpha$ and the larger exchange constant
$J_1$ (or equivalently to $\delta$ and $J$).  The ratio
$\Delta(\alpha)/J_1$ for the $S = 1/2$ AF alternating-exchange Heisenberg
chain was computed to high ($\leq 1\%$) accuracy for $0 \leq \alpha \leq
0.9$, in $\alpha$ increments of 0.1, using multiprecision methods by
Barnes, Riera and Tennant.\cite{Barnes1998}  They found that their
calculations could be parametrized very well by the simple expression
\begin{mathletters}
\label{EqDimParams2:all}
\begin{equation}
\frac{\Delta(\alpha)}{J_1} \approx (1 -
\alpha)^{3/4}(1 + \alpha)^{1/4}~,\label{EqDimParams2:a}
\end{equation} or equivalently
\begin{equation}
\frac{\Delta(\delta)}{J} \approx 2\,\delta^{3/4}~.
\label{EqDimParams2:b}
\end{equation}
\end{mathletters} 

An expression for $\Delta(\delta)/J$ which is thought to be more accurate
($\pm 0.0002$) over the entire range $0 \leq \delta \leq 1$, obtained by
fitting numerical $\Delta(\delta)/J$ data by a generalized form of
Eq.~(\ref{EqDimParams2:b}), is\cite{Johnston1999}
\begin{mathletters}
\label{EqD(d):all}
\begin{equation} {\Delta(\delta)\over J} = 2\,\delta\,^{y(\delta)}~,
\label{EqD(d):a}
\end{equation} where the exponent $y(\delta)$ is given by
\begin{eqnarray} y(\delta) = y(1) &+& n_1\tanh\bigg[{\ln\delta\over m_1}
\ln\Big({\ln\delta\over m_2}\Big)\bigg]\nonumber\\ &+&
n_2\tanh^2\bigg[{\ln\delta\over m_1}\ln\Big({\ln\delta\over
m_2}\Big)\bigg]
\label{EqD(d):b}
\end{eqnarray} with parameters
\[ y(1) = 0.74922~,~~~n_1 = 0.00776~,~~~n_2 = -0.00685~,
\]
\begin{equation} m_1 = 3.3297~,~~~m_2 = -2.2114~.
\label{EqD(d):c}
\end{equation}
\end{mathletters}
The expression for $\Delta(\delta)/J$ in Eqs.~(\ref{EqD(d):all}) can be
transformed into an expression for $\Delta(\alpha)/J_1$ using the
conversion expressions~(\ref{EqDimParams:all}).

For notational convenience, we define the reduced spin susceptibility
$\chi^*$, reduced temperature $t$ and reduced spin gap $\Delta^*$ as
\begin{equation}
\chi^* \equiv \frac{\chi^{\rm spin} J_1}{Ng^2\mu_{\rm B}^2}~,~~~t \equiv
{k_{\rm B}T\over J_1}~,~~~\Delta^* = {\Delta\over J_1}~,
\end{equation} where $\chi^{\rm spin}$ is the spin susceptibility, $N$ is
the number of spins, $k_{\rm B}$ is Boltzmann's constant and $\chi^*$
depends on both $t$ and the alternating exchange parameter $\alpha \equiv
J_2/J_1$.

An accurate but unwieldy two-dimensional function $\chi^*(t,\alpha)$ for
the $S = 1/2$ AF alternating-exchange Heisenberg chain has been derived
for the entire range \mbox{$0 \leq \alpha \leq 1$} of the alternation
parameter by a global fit to numerical quantum Monte Carlo
(QMC) simulations and transfer-matrix density-matrix renormalization group
(TMRG) and Bethe ansatz $\chi^*(t,\alpha)$
calculations,\cite{Johnston1999} which we will not reproduce here but
will explicitly use to model $\chi(T)$ data for both AP- and HP-${\rm
(VO)_2P_2O_7}$.  The absolute accuracy of this function for $0 \leq
\alpha \leq 1$ and $0.01 \lesssim t$ is estimated to be
$\lesssim 2\times 10^{-4}$, which corresponds to $\lesssim 0.1$\,\% of the
susceptibility at the broad maximum.  For practical purposes of fitting
experimental $\chi(T)$ data, this function can be considered to be exact
for $t \gtrsim 0.01$.

Troyer, Tsunetsugu, and W\"{u}rtz\cite{Troyer1994} have derived a general 
expression for the low-$T$ limit of $\chi^*(t)$ for a one-dimensional spin
system with a spin gap, assuming that (i)~the one-magnon dispersion
relation is nondegenerate (apart from the Zeeman degeneracy), (ii)~the
lowest magnetic excited states are one-magnon $S = 1$ triplet
excitations, (iii)~$k_{\rm B}T\ll \Delta$ {\em and} $k_{\rm B}T \ll$
one-magnon bandwidth, and (iv)~the one-magnon dispersion relation $E(k)$
is parabolic near the minimum according to (in the present notation)
\begin{equation}
\varepsilon_k \equiv {E(k)\over J_1} = \Delta^* +  c^*(ka)^2~,
\label{EqEps(k)}
\end{equation}
where $k$ is the wavevector in the direction of the 1D system and $a$
is the (average) nearest-neighbor spin-spin distance.  These assumptions
hold for the present case of the $S = 1/2$ AF alternating-exchange
Heisenberg chain except for the limit $\alpha = 0$ as
discussed below and for $\alpha = 1$ for which
$\Delta = 0$.\cite{Johnston1999}  With these four assumptions,
$\chi^*(t)$ is given by\cite{Troyer1994}
\begin{mathletters}
\label{EqTroyer:all}
\begin{equation}
\chi^*(t) = {A\over \sqrt{t}}\,{\rm e}^{-\Delta^*/t}~,~~~(t\ll
\Delta^*,\,{\rm bandwidth}/J_1)
\label{EqTroyer:a}
\end{equation}
with
\begin{equation}
A = {1\over 2\sqrt{\pi c^*}}~.
\label{EqTroyer:b}
\end{equation}
\end{mathletters}

The dimensionless dispersion parameter $c^*$ in Eq.~(\ref{EqEps(k)}) has a
unique relationship to the reduced spin gap $\Delta^*$ for any given 1D
spin system.  For example, according to the model of
Ref.~\onlinecite{Johnston1999}, this relationship for the $S = 1/2$ AF
alternating-exchange Heisenberg chain gives the value of the parameter
$A$ in Eq.~(\ref{EqTroyer:a}) as
\begin{equation}
A = {\sqrt{\Delta^*}\over \sqrt{2\pi}f(\Delta^*)}~,
\label{EqA}
\end{equation}
which in turn yields the low-$T$ limit of $\chi^*(t)$ in
Eq.~(\ref{EqTroyer:a}) as
\begin{mathletters}
\label{EqMyLoTChi:all}
\begin{equation}
\chi^*(t) = {1\over\sqrt{2\pi}\,f(\Delta^*)}\,\Big({\Delta^*\over
t}\Big)^{1/2} {\rm e}^{-\Delta^*/t}~,~~~(t\ll \Delta^*)
\label{EqMyLoTChi:a}
\end{equation}
where the dimensionless function $f(\Delta^*)$ is the solution of
\begin{equation}
{\rm E}\bigg[-{f^2(\Delta^*)\over{\Delta^*}^2}\bigg] =
{\pi\over 2\Delta^*}
\label{EqMyLoTChi:b}
\end{equation}
\end{mathletters}
and E($x$) is the complete elliptic integral of the second kind.

From Eq.~(\ref{EqA}), the parameter $A$ in Eq.~(\ref{EqTroyer:a}) is
not an independently adjustable parameter but instead is a unique
function of the reduced spin gap $\Delta^*$,\cite{Johnston1999} as was
also previously inferred for two-leg spin ladders.\cite{Johnston1996}  In
addition, we see from Eq.~(\ref{EqMyLoTChi:a}) that the two independent
parameters of $\chi^*(t,\alpha)$ can be written for low temperatures as
$\chi^*(t/\Delta^*,\Delta^*)\equiv \chi^*(k_{\rm B}T/\Delta,\Delta/J_1)$. 
Finally, and perhaps most importantly, the high-temperature limit of the
low-temperature regime in which $\chi^*(t)$ is closely approximated by
Eqs.~(\ref{EqMyLoTChi:all}) is of order
$\Delta^*/10$.\cite{Johnston1999}  At such low temperatures
$\chi(T)$ is immeasurably small, and hence the spin gap obtained
by analyzing experimental $\chi(T)$ data for various compounds up to
temperatures corresponding to a sizable fraction of $\Delta^*$ using
Eq.~(\ref{EqTroyer:a}) leads to fitted spin gap values which may be
significantly different from the actual spin gaps.

The result for the low-$t$ limit of $\chi^*(t)$ in
Eq.~(\ref{EqTroyer:a}) is not valid for the isolated dimer, which is one
limit of the alternating-exchange chain with \mbox{$\alpha = J_2 = 0$},
because the assumption \,(iii) \,that $k_{\rm B}T\ll$ one-magnon bandwidth
required for that equation to hold is violated at all finite
temperatures.  For the isolated dimer, the one-magnon bandwidth is
identically zero and the reduced spin gap is $\Delta^* = 1$.  The reduced
spin susceptibility is given exactly by
\begin{mathletters}
\label{EqDimer:all}
\begin{equation}
\chi^{*,{\rm dimer}}(t) = {1\over t(3 + {\rm e}^{1/t})}
\label{EqDimer:a}
\end{equation}
with the low-temperature limit
\begin{equation}
\chi^{*,{\rm dimer}}(t) = {1\over t}\,{\rm e}^{-1/t}~~~(t \ll 1)~.
\label{EqDimer:b}
\end{equation}
\end{mathletters}
The temperature dependence of the prefactor to the exponential term in
Eq.~(\ref{EqDimer:b}) is different from that in Eq.~(\ref{EqTroyer:a}). 
As discussed in Ref.~\onlinecite{Johnston1999}, a crossover occurs with
decreasing temperature at low temperatures in the effective prefactor from
a $1/t$ dependence to a $1/\sqrt{t}$ dependence if $0 < \alpha \ll 1$.

As noted above, the form for the low-$T$ behavior of the spin
susceptibility in Eq.~(\ref{EqTroyer:a}) is valid only at very low
temperatures.  Many years ago, Bulaevskii found that his numerical
values of $\chi^*(t,\alpha)$ for the $S = 1/2$ AF alternating-exchange
Heisenberg chain, computed from an analytic theory based on the
Hartree-Fock approximation, could be fitted over a relatively large
temperature interval $0.033 \leq t \leq 1/4$ by\cite{Bulaevskii1969}
\begin{equation}
\chi^*(t) = {A\over t}\,{\rm e}^{-\Delta^*/t}~,
\label{EqBul}
\end{equation}
and he tablulated $A$ and $\Delta^*$ versus the alternation parameter
$\alpha$.  A recent extensive numerical study\cite{Johnston1999} of his
theory confirmed that the numerical predictions of his theory in the
above-cited low-temperature range are fitted better by the
form~(\ref{EqBul}) than by~(\ref{EqTroyer:a}).  In addition, this study
showed that although the fitting parameter $\Delta^*(\alpha)$
approximately follows the actual spin gap of \mbox{Bulaevskii's} theory,
significant discrepancies occur.\cite{Johnston1999}  Finally, a detailed
numerical comparison of the prediction of Bulaevskii's theory for
$\chi^*(t,\alpha)$ with QMC simulations and TMRG calculations of this
quantity showed that \mbox{Bulaevskii's} theory is unsuitable for
accurately extracting  $\alpha$ values from experimental $\chi(T)$ data
when $\alpha \lesssim 1$.\cite{Johnston1999}

For Heisenberg spin lattices consisting of identical spin subsystems
with susceptibility $\chi^*_0(t)$ which are weakly coupled to each
other, the molecular field theory (MFT) prediction for the reduced spin
susceptibility $\chi^*(t)$ in the paramagnetic state of the system
is
\begin{mathletters}
\label{EqMFT:all}
\begin{equation}
\chi^*(t) = \frac{\chi^*_0(t)}{1 + \lambda\,\chi^*_0(t)}~,
\label{EqMFT:a}
\end{equation} 
or equivalently
\begin{equation} {1\over \chi^*(t)} = {1\over \chi^*_0(t)} + \lambda~,
\label{EqMFT:b}
\end{equation} where the MFT coupling constant $\lambda$ is given by
\begin{equation}
\lambda = \sum_j^{\ \ \ \ \ \ \prime} {J_{ij}\over J^{\rm max}}~,
\label{EqMFT:c}
\end{equation}
\end{mathletters} the prime on the sum over $j$ signifies that the sum is
only taken over exchange bonds $J_{ij}$ from a given spin $\bbox{S}_i$ to
spins $\bbox{S}_j$ not in the same spin subsystem, and $J^{\rm max}$ is the
exchange constant in the system with the largest magnitude.  By definition,
the expression for $\chi^*_0(t)$ does not contain any of these $J_{ij}$
interactions which are external to a subsystem.  Within MFT,
Eqs.~(\ref{EqMFT:all}) are correct at each temperature in the
paramagnetic state not only for bipartite AF spin systems, but also for
any system containing subsystems coupled together by any set of FM and/or
AF Heisenberg exchange interactions.

\section{Modeling of Experimental $\bbox{\chi(T)}$ Data}
\label{SecEval}

\subsection{Introduction}
\label{SecExpIntro}

We fitted the $\chi(T)$ data per mole of V spins-1/2 in 
(VO)$_2$P$_2$O$_7$ by the general expression
\begin{mathletters}
\label{EqChiExp:all}
\begin{equation}
\chi(T) = \chi_0 + \frac{C_{\rm imp}}{T - \theta_{\rm imp}} + \chi^{\rm
spin}(T)~,
\label{EqChiExp:a}
\end{equation} with
\begin{equation}
\chi_0 = \chi^{\rm core} + \chi^{\rm VV}
\label{EqChiExp:b}
\end{equation} and
\begin{eqnarray}
\chi^{\rm spin}(T) &=& {N_{\rm A}g^2\mu_{\rm B}^2\over J_1}
\chi^*(t)\nonumber\\ &=& \Big(0.3751\,\frac{\rm cm^3\,K}{\rm mol}\Big)
{g^2\over J_1/k_{\rm B}}\chi^*\Big({k_{\rm B}T\over J_1}\Big)~,
\label{EqChiExp:c}
\end{eqnarray}
\end{mathletters}  where $N_{\rm A}$ is Avogadro's number, $\mu_{\rm B}$ is the Bohr
magneton, $k_{\rm B}$ is Boltzmann's constant and $g$ is the spectroscopic
splitting factor ($g$ factor) appropriate to a particular direction of the
applied magnetic field with respect to the crystal axes.  

The first term $\chi_0$ in Eq.~(\ref{EqChiExp:a}), according to
Eq.~(\ref{EqChiExp:b}), is the sum of the nearly isotropic orbital
diamagnetic atomic core contribution $\chi^{\rm core}$ and the 
anisotropic orbital paramagnetic Van Vleck contribution
$\chi^{\rm VV}$ which are normally nearly independent of $T$.  Using the
values $\chi^{\rm core} = -8,\ -12$ and $-47\times 10^{-6}$ cm$^3$/mol
for V$^{+4}$, O$^{-2}$ and ${\rm (PO_4)^{-3}}$,
respectively,\cite{Landolt1979} we obtain
\begin{equation}
\chi^{\rm core} = -6.1\times 10^{-5} {\rm {cm^3\over mol\,V}}
\label{EqChiCore}
\end{equation}
for \mbox{(VO)$_2$P$_2$O$_7$}.  The second term in Eq.~(\ref{EqChiExp:a})
is an extrinsic impurity Curie-Weiss term with impurity Curie constant
$C_{\rm imp}$ and Weiss temperature $\theta_{\rm imp}$ which gives rise to
a low-temperature upturn in $\chi(T)$ which is not predicted by theory for
the third term, the intrinsic spin susceptibility $\chi^{\rm spin}(T)$,
and is assumed to arise from finite chain segments containing an odd
number of spins, impurity phase intergrowths in the crystals,
paramagnetic impurity phases  and/or defects.  The $C_{\rm imp}$ and
$\theta_{\rm imp}$ parameters can be anisotropic if the paramagnetic
impurity principal directions are fixed with respect to the crystal axes,
as can occur in a single crystal of a material such as studied here in
Sec.~\ref{SecXtalChi}, rather than being randomly distributed.

Unless otherwise stated, we assume that the spin susceptibility
$\chi^{\rm spin}(T)$ in Eq.~(\ref{EqChiExp:a}), written in terms of
$\chi^*(t,\alpha)$ in Eq.~(\ref{EqChiExp:c}), is the intrinsic spin
susceptibility per mole of spins-1/2 in an AF alternating-exchange
Heisenberg chain.  The explicit expression for $\chi^*(t,\alpha)$ of this 
chain is given in Ref.~\onlinecite{Johnston1999}. 
In Eq.~(\ref{EqChiExp:c}), $J_1$ is the larger of the two ($J_1$ and $J_2$
with $J_1 > J_2 > 0$) AF alternating exchange constants along the 
alternating-exchange chain, as denoted in Sec.~\ref{SecThy} above.

One of the parameters entering the calculated spin susceptibility
$\chi^{\rm spin}(T)$ in Eq.~(\ref{EqChiExp:c}) is the $g$ value of the
V magnetic moments.  Measurements of the anisotropic $g$ values of the
spins in both AP-(VO)$_2$P$_2$O$_7$ and \mbox{HP-(VO)$_2$P$_2$O$_7$} have
been carried out using ESR measurements\cite{Schwenk1996,Prokofiev1998}
and the results are listed in Table~\ref{TabESR}.  The significant
differences between the $g$ values of the two phases of ${\rm
(VO)_2P_2O_7}$ reflect the differences in the local bonding of the V
atoms with the coordinating O atoms in the two structures.  Comparison of
the average $g$ value for HP-${\rm (VO)_2P_2O_7}$ with those for V in the
``trellis layer'' compounds $R{\rm V_2O_5}$ with $R$ = Ca, Mg and Na, 
\cite{Onoda1996,Onoda1998,Vasilev1997,Schmidt1998,Lohmann1997,Onoda1999}
also shown in Table~\ref{TabESR}, suggests that the local crystalline
electric field (CEF) at the V sites in HP-${\rm (VO)_2P_2O_7}$ is closer
to that in these compounds than to the CEF in AP-${\rm (VO)_2P_2O_7}$.

A measure of the goodness of a fit to experimental \ $\chi(T)$ \ data \ is
\ the \ statistical \ $\chi^2$ \ per \ degree \ of \ freedom 

\vglue0.43in
% Table II
\begin{table}
\caption{$g$-factors parallel ($g_{||}$) and perpendicular ($g_{\bot}$) to
the principal local crystalline electric field and/or crystal structure
axis and the powder-averaged value $g = \sqrt{(g_1^2 + g_2^2 +
g_3^2)/3}$ for V$^{+4}$ $S = 1/2$ species in several vanadium
oxide compounds.  Samples are polycrystalline unless otherwise noted. 
The literature references are given in the last column.}
\begin{tabular}{lcccr} Compound & $g_{||}$ & $g_{\bot}$ & $g$ & Ref.\\
\hline 
AP-${\rm (VO)_2P_2O_7}$& 1.94 & 1.98 & 1.97 & \onlinecite{Schwenk1996}\\
~~crystal   & 1.937($a$) & 1.984($b,c$) & 1.969 &
\onlinecite{Prokofiev1998}\\ HP-${\rm (VO)_2P_2O_7}$ & 1.928(1)($a$) &
1.974(1)($b$), & 1.958 &
\onlinecite{Saito2000}\\ 
~~(crystal)&&1.971(1)($c$)\\
${\rm CaV_2O_5}$ &&& 1.957(1) & \onlinecite{Onoda1996} \\ MgV$_2$O$_5$
&      &      & 1.96 & \onlinecite{Onoda1998}\\
NaV$_2$O$_5$ (crystal) & 1.938(2) & 1.972(2) & 1.961(2) &
\onlinecite{Vasilev1997}\\ ~~crystal & 1.936(2) &  &  &
\onlinecite{Schmidt1998}\\ ~~crystal & 1.95 & 1.97 & 1.96 &
\onlinecite{Lohmann1997}\\ ~~crystal & 1.936 & 1.974, 1.977 & 1.962 &
\onlinecite{Onoda1999} \\
\end{tabular}
\label{TabESR}
\end{table}
% Figure 4
\begin{figure}
\epsfxsize=3.4in
\centerline{\epsfbox{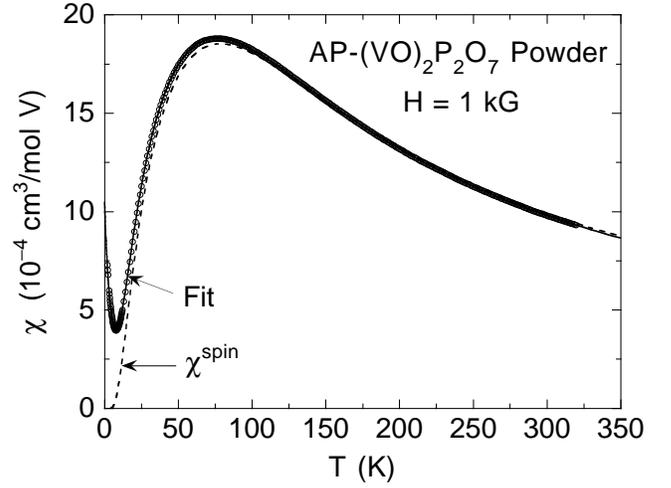}}
\vglue 0.1in
\caption{Magnetic susceptibility $\chi$ versus temperature $T$ for
a powder sample of AP-(VO)$_2$P$_2$O$_7$ ($\circ$) from 2\,K to
350\,K\@.  The solid curve is a two-dimensional fit to the 357 data points
using Eq.~(\protect\ref{EqChiExp:a}) assuming a spin susceptibility
$\chi^{\rm spin}(T)$ consisting of those of equal numbers of two
independent isolated $S = 1/2$ antiferromagnetic alternating-exchange
Heisenberg chains.  The dashed curve is the fitted $\chi^{\rm spin}(T)$. 
The fitted  exchange constants and the derived spin gaps for the two
chains are listed in the column labeled ``Fit~6'' in
Table~\protect\ref{TabFitAPVOPOPowder}.}
\label{Yama1kGChiPoly2ChnFit}
\end{figure}

\noindent
\begin{equation}
{\chi^2\over{\rm DOF}} \equiv {1\over N_{\rm p} - P}\sum_{i=1}^{N_{\rm p}}
[\chi(T_i) - {\chi_{\rm fit}}(T_i)]^2~,
\end{equation}
where $N_{\rm p}$ is the number of data points in the data set and $P$ is
the number of independent fitting parameters.  This is the quantity that
is minimized during our nonlinear least-squares fits to experimental
$\chi(T)$ data.  An additional measure of
the quality of a fit is the relative rms deviation $\sigma_{\rm rms}$ of
the fit from the data, given by
\begin{equation}
\sigma_{\rm rms}^2 \equiv {1\over N_{\rm p}}\sum_{i=1}^{N_{\rm p}}
{\left[{\chi(T_i) - \chi_{\rm fit}(T_i)\over \chi(T_i)}\right]^2}~.
\end{equation}
The fits were all carried out on a 400\,MHz
Macintosh\,G3 computer using the software Mathematica\,3.0.

\subsection{Magnetic Susceptibility of
AP-(VO)$_{\bbox{2}}$P$_{\bbox{2}}$O$_{\bbox{7}}$}
\label{SecPowderChiAPVOPO}
\vglue0.13in

\subsubsection{Powder sample}
\label{SecAPVOPOPowderChi}

The purpose of the present section is to test consistency with
experimental $\chi(T)$ data of the model of \mbox{Yamauchi} and
co-workers for AP-(VO)$_2$P$_2$O$_7$,\cite{Yamauchi1999,Kikuchi1999}
discussed in the Introduction, in which this compound is proposed
to consist magnetically of equal numbers of two independent types of
isolated $S = 1/2$ AF alternating-exchange Heisenberg chains with spin
gaps of about 35 and 68\,K, respectively.

The $\chi(T)$ of a polycrystalline (``powder'') sample of
AP-(VO)$_2$P$_2$O$_7$ of mass 172.2\,mg and with a moss-green color was
measured from 2 to 350\,K in a magnetic field of 1~kG and the results are
shown as the open circles in Fig.~\ref{Yama1kGChiPoly2ChnFit}.  The
details of the sample preparation will be presented elsewhere.  The color
of the sample indicates that it is stoichiometric with a vanadium
oxidation state very close to +4.\cite{Prokofiev1998}  The sample was a
cylinder of 4\,mm diameter and 7\,mm length.  There was no difference
between field-cooled and zero-field-cooled $\chi(T)$ measurements.  The
quality of the sample, judging from the very small Curie-Weiss upturn at
low temperatures due to magnetic impurities and/or defects, is better than
previously reported for any powder sample and is about the same as
recently reported for a high-quality single crystal\cite{Prokofiev1998}
as shown in Figs.~\ref{AP-VOPOXtals}
and~\ref{AP-VOPOChiabcFit}(b) below. 

We will describe in detail our modeling results for this sample to
indicate how the parameters and the quality of fit change for various
types of fits.  Similar variations were found from modeling $\chi(T)$ 
data for two \mbox{AP-(VO)$_2$P$_2$O$_7$} single crystals below.  To make
contact with previous modeling of $\chi(T)$ for this material, we first
fitted the data by Eqs.~(\ref{EqChiExp:all}) assuming that $\chi^{\rm
spin}(T)$ is due to a single type of $S = 1/2$ AF alternating-exchange
Heisenberg chain, where the $g$ value is either fixed at the
powder-averaged value $g = 1.969$ determined from ESR measurements for
AP-(VO)$_2$P$_2$O$_7$ as shown in Table~\ref{TabESR}, yielding ``Fit~1'',
or allowed to vary during the fit (``Fit~2'').  Throughout 
\widetext
% Table III
\begin{table}
\caption{Fitted and derived parameters for $\chi(T)$ of a high-quality
powder sample of AP-(VO)$_2$P$_2$O$_7$.  A derived quantity is marked by
an asterisk (*).  A quantity with a ``$\equiv$'' in front
of it was constrained to have the value listed and was not fitted. 
Quantities $A^{(1)}$ and $A^{(2)}$ are quantities associated with two
independent isolated chains, respectively.  The $g$~value in Fits~1, 3,
and~5 were constrained to be the powder-averaged value 1.969 from ESR
measurements (see Table~\protect\ref{TabESR}), whereas in Fits~2, 4,
and~6 the $g$ value was fitted.  Fits~1 and~2 assume a single type of
isolated alternating-exchange chain, whereas Fits~3--6 assume the
presence of two independent isolated alternating-exchange chains.  In
Fits~3 and~4 the two respective spin gaps were constrained to have the
values found (Ref.~\protect\onlinecite{Garrett1997a}) from inelastic
neutron scattering measurements, whereas in Fits~5 and~6 the two spin
gaps were each allowed to vary independently.  Our favored fit parameters
are those of Fit~6.}
\begin{tabular}{lcccccc}
Quantity & Fit 1 & Fit 2 & Fit 3 & Fit 4 & Fit 5 & Fit 6\\
\hline
$\chi_0\ \big({\rm 10^{-5}{cm^3\over mol\,V}}\big)$ & $-$1.8(1) &
$-$4.5(6) & $-$1.37(6) & $-$5.2(1) & $-$1.99(2) & $-$2.3(1) \\
$C_{\rm imp}\ \big({\rm 10^{-3}\,{cm^3\,K\over mol\,V}}\big)$ & 5.8(1)
& 6.3(2) & 3.72(2) & 4.52(3) & 4.26(2) & 4.29(2) \\
$\theta_{\rm imp}$ (K) & $-$6.4(2) & $-$6.9(3) & $-$3.08(4) & $-$3.95(3) &
$-$3.89(2) & $-$3.91(2) \\
$g$ & $\equiv$ 1.969 & 1.993(6) & $\equiv$ 1.969 & 2.006(1) & $\equiv$
1.969 & 1.972(1) \\
$J_1^{(1)}/k_{\rm B}$ (K) & 130.9(1) & 132.4(4) & 137.0(8)* &
144.4(4)* & 128.7(4) & 130.3(7) \\
$J_1^{(2)}/k_{\rm B}$ (K) &---&---& 122.7(6)* & 126.0(4)* &
129.9(4) & 128.8(5) \\ 
$J_2^{(1)}/k_{\rm B}$ (K) & 97.9(2)* & 99.8(5)* & 118.2(6)* &
121.7(3)* & 107.6(5)* & 109.5(9)* \\
$J_2^{(2)}/k_{\rm B}$ (K) &---&---& 76.4(5)* & 75.2(2)* &
83.7(4)* & 82.6(6)* \\ 
$\alpha^{(1)}$ & 0.748(1) & 0.754(2) & 0.863(1)* & 0.8728(5)* &
0.836(1) & 0.840(2) \\
$\alpha^{(2)}$ &---&---& 0.622(2)* & 0.6178(7)* & 0.644(1) &
0.641(2) \\
$J^{(1)}/k_{\rm B}$ (K) & 114.4(2)* & 116.1(5)* & 127.6(6)* &
136.2(4)* & 118.1(4)* & 119.9(8)* \\
$J^{(2)}/k_{\rm B}$ (K) &---&---& 99.6(5)* & 98.5(2)* &
106.8(4)* & 105.7(6)* \\ 
$\delta^{(1)}$ & 0.1442(6)* & 0.140(1)* & 0.0735(5) & 0.0679(3) &
0.0893(6)* & 0.087(1)* \\
$\delta^{(2)}$ &---&---& 0.233(1) & 0.2363(5) & 0.2165(7)* &
0.219(2)* \\
$\Delta^{(1)}/k_{\rm B}$ (K) & 53.6(3)* & 53.2(4)* & $\equiv$ 36.2 &
$\equiv$ 36.2 & 38.8(3)* & 38.6(5)* \\
$\Delta^{(2)}/k_{\rm B}$ (K) &---&---& $\equiv$ 66.7 & $\equiv$ 66.7 &
67.7(4)* & 67.5(5)* \\
${\chi^2\over {\rm DOF}}$ $\big({\rm 10^{-5}{cm^3\over mol\,V}}\big)^2$
& 1.31 & 1.24 & 0.194 & 0.0379 & 0.0130 & 0.0117 \\
$\sigma_{\rm rms}$ (\%) & 1.97 & 1.94 & 0.617 & 0.226 & 0.138 & 0.132 \\
\end{tabular}
\label{TabFitAPVOPOPowder}
\end{table}

\narrowtext
\noindent the modeling
in this section, we use the expression~(\ref{EqD(d):all}) to determine
the spin gap from the fitted exchange constants for an
alternating-exchange chain.  The parameters obtained from each fit are
shown in Table~\ref{TabFitAPVOPOPowder}, together with the statistical
$\chi^2$/DOF and $\sigma_{\rm rms}$ for each fit.  The defect and/or
impurity Curie constant is equivalent to the contribution of about
1.6\,mol\,\% with respect to~V of spins~1/2 \ with \ $g = 2$. \  The
\ values \ of \ the \ alternating exchange constants $J_1$ and $J_2$ and the spin gap
$\Delta$ are respectively similar in each fit and are about the same as
previously estimated from similar fits to $\chi(T)$ data for this
compound, respectively (see Table~\ref{TabJAPVOPO}).  Next, we fitted the
$\chi(T)$ data using the above model of
\mbox{Yamauchi} and co-workers for
\mbox{AP-(VO)$_2$P$_2$O$_7$},\cite{Yamauchi1999,Kikuchi1999} but where we
constrained the spin gaps to be 36.2 and 66.7\,K as found from the
inelastic neutron scattering measurements,\cite{Garrett1997a} and again
either fixed $g = 1.969$ (``Fit~3'') or allowed $g$ to vary during the
fit (``Fit~4'').   In order to enforce the constraint on the two spin
gaps using the expression for $\Delta(\delta)/J$ in
Eqs.~(\ref{EqD(d):all}), it was more convenient to use $J$ and $\delta$
as the independent parameters in $\chi^{\rm spin}(T)$ during our
least-squares fits instead of $J_1$ and $\alpha$.  The parameters obtained
from the two fits are shown in Table~\ref{TabFitAPVOPOPowder}, together
with other parameters derived from the fitted ones.  As can be seen
from the values of the $\chi^2/$DOF and $\sigma_{\rm rms}$, the qualities
of the two fits show dramatic improvements over those of Fits~1 and~2
where only a single type of alternating-exchange chain was assumed in the
model-  

\newpage
\noindent ing.  However, the $g$ value obtained from Fit~4 is
somewhat larger than expected.

Finally, we fitted the same data set in Fig.~\ref{Yama1kGChiPoly2ChnFit}
using the above model of \mbox{Yamauchi} and co-workers for
\mbox{AP-(VO)$_2$P$_2$O$_7$},\cite{Yamauchi1999,Kikuchi1999} where we
again either fixed $g = 1.969$ (``Fit~5'') or allowed $g$ to vary during
the fit (``Fit~6''), but where we did not constrain the fitting parameters
$J_1$ and $\alpha$ of the two independent chains to yield the
respective spin gaps found from the inelastic neutron scattering
measurements.  The fitted parameters are listed for each fit in
Table~\ref{TabFitAPVOPOPowder}, together with the statistical
$\chi^2$/DOF and $\sigma_{\rm rms}$ and the derived $\Delta$ for each
fit.  We checked that the identical fitted parameters are obtained
for Fit~5 independent of whether the starting parameters are the fitted
parameters of Fit~1, for which the exchange constants in the two chains
are identical, or of Fit~3 for which they are different.  Fit~6 is shown
as the solid curve in Fig.~\ref{Yama1kGChiPoly2ChnFit} and the fitted
$\chi^{\rm spin}(T)$ is shown as the dashed curve.

The fitted and derived parameters for Fits~5 and~6 in
Table~\ref{TabFitAPVOPOPowder} exhibit a number of important features. 
First, the qualities of Fits~5 and~6 to the data are far superior to
those of Fits~1 and~2.  Second, the values of the alternation parameters
and spin gaps for the two independent isolated chains of the model did not
converge to the same respective values for the two chains, but rather are
clearly differentiated.  Third, the fitted $g$ value from Fit~6 is
identical within the respective errors with the powder averaged $g$ value 
in Table~\ref{TabESR} determined from ESR measurements.  Fourth, and
perhaps most importantly, the two spin gaps derived from the respective
exchange constants for the two chains are respectively nearly identical
to the two spin gaps found from the single-crystal inelastic neutron
scattering measurements by Garrett~{\it et al.}\cite{Garrett1997a} and
with the values inferred previously by Yamauchi and co-workers from
high-field magnetization measurements and a subset of the NMR
measurements.\cite{Yamauchi1999,Kikuchi1999}  It seems very unlikely that
the two spin gaps we deduce from this model could be so close to those
determined from other independent measurements without the model being
essentially correct.  The exchange constants and spin gaps we derived
from $\chi(T)$ data for the same powder sample in $H = 50$\,kG, data
which are not otherwise discussed here, are identical within the
respective errors to those we obtained above for $H = 1$\,kG\@.  Finally,
we will see in Sec~\ref{SecCrushedXtalChi} below that when the two-chain
model is used to extract the exchange constants within the proposed
single-chain high-pressure phase HP-(VO)$_2$P$_2$O$_7$, essentially the
same exchange constants and spin gaps are obtained for both chains of the
model.  This result indicates that our fitting procedure can clearly
differentiate between pairs of chains that have the same or different
spin gaps, respectively.

We conclude that our analysis of $\chi(T)$ is precisely consistent
with the model of Yamauchi and co-workers for the
nature of the important spin interactions in AP-(VO)$_2$P$_2$O$_7$.  Our
values of the spin gaps of the two independent isolated
alternating-exchange chains of the model are in good agreement with those
determined from their high-field magnetization and NMR
measurements\cite{Yamauchi1999,Kikuchi1999} and with the two values
determined from the inelastic neutron scattering
measurements,\cite{Garrett1997a} respectively.

\subsubsection{MFT analysis of interchain coupling}
\label{SecAP-VOPOMFT}
\vglue-0.02in
The neutron scattering measurements on single crystals showed
unambiguously that interchain coupling $J_a$ along the $a$ axis of the
structure, perpendicular to the alternating-exchange chains and parallel
to the legs of the structural two-leg ladders (see
Fig.~\ref{VOPOStruct2}), is not negligible.\cite{Garrett1997a}  However
the ratio $|J_a/J_1|$ was estimated from fits to the data to be only
2--3\,\%, where $J_1$ is the larger of the two exchange couplings along
the alternating-exchange chains running along the
$c$ axis.\cite{Garrett1997a}  Another estimate can be obtained as one-half
the ratio of the average total dispersion of the two presumed one-magnon
bands in the $a$ axis direction [16(4)\,K] to the one-magnon excitation
energy along the direction of the alternating-exchange chains at the zone
boundary (180\,K), yielding a slightly larger $|J_a/J_1|\approx
4.5(10)$\,\%.  This interchain coupling along the $a$ axis was of course
ignored in the fits to the experimental $\chi(T)$ in the previous 
section.  Here we obtain an estimate of the strength of this interchain
coupling $J_a$ from analysis of the powder $\chi(T)$ data presented in the
previous section.  In the absence of accurate calculations of $\chi^*(t)$
for this case, we will utilize the prediction of MFT given in
Eqs.~(\ref{EqMFT:all}) for the influence of the interchain coupling on
$\chi(T)$.

In order to apply Eqs.~(\ref{EqMFT:all}) to the present modeling
framework in which two distinct types of alternating-exchange chains~(1)
and~(2) are assumed to be present, one must appropriately define the
``isolated subsystem'' discussed in Sec.~\ref{SecThy}.  Here, an isolated
subsystem consists of one set of the two chains~(1) and~(2). 
Thus, the reduced susceptibility of our isolated subsystem is
\begin{eqnarray}
\chi_0^* &\equiv& {\chi J_1^{\rm max}\over Ng^2\mu_{\rm B}^2} \nonumber\\
&=& {1\over 2}\left[{J_1^{\rm max}\over J_1^{(1)}}\,\chi^*_{\rm
chain}(t^{(1)},\alpha^{(1)}) + {J_1^{\rm max}\over J_1^{(2)}}\,\chi^*_{\rm
chain}(t^{(2)},\alpha^{(2)})\right]\nonumber
\end{eqnarray}
where
\begin{equation}
t^{(1)} \equiv {k_{\rm B}T\over J_1^{(1)}},~~t^{(2)} \equiv {k_{\rm
B}T\over J_1^{(2)}},~~J_1^{\rm max} \equiv
\max\Big[J_1^{(1)},J_1^{(2)}\Big]~.
\label{EqMFTChi*}
\end{equation}
Then the MFT coupling constant $\lambda$ in Eqs.~(\ref{EqMFT:all}) is,
according to Eq.~(\ref{EqMFT:c}), the average interchain coupling in the
$a$ axis direction of a spin in one of the two distinct
alternating-exchange chains with all spins in the respective adjacent
alternating-exchange chains.  Assuming \,that \,a \,spin \,in \,each chain
is coupled to two nearest 
% Figure 5
\begin{figure}
\epsfxsize=3in
\centerline{\epsfbox{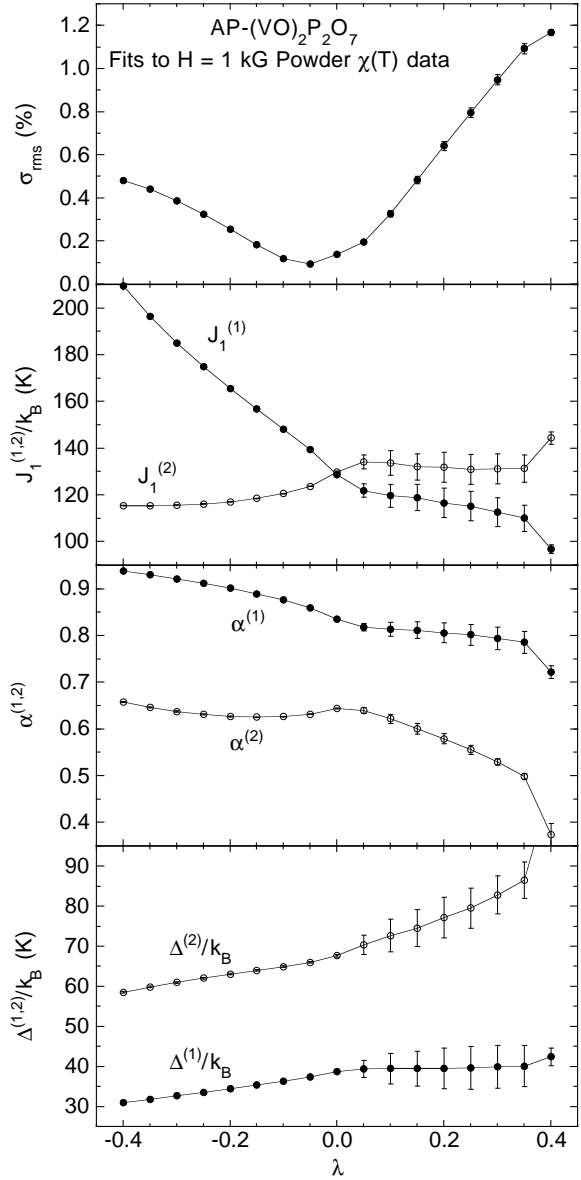}}
\vglue 0.1in
\caption{Parameters $J_1^{(1,2)}$ and $\alpha^{(1,2)}$ and rms deviation
$\sigma_{\rm rms}$ of the fits to the powder $\chi(T)$ data for
AP-(VO)$_2$P$_2$O$_7$ in Fig.~\protect\ref{Yama1kGChiPoly2ChnFit} by MFT
for coupled alternating-exchange chains of types~(1) and~(2) versus the 
MFT interchain coupling constant $\lambda$.  A $g = 1.969$ was assumed
in all of the fits.  The spin gaps $\Delta^{(1,2)}$ for the two distinct
chains~(1) and~(2), derived from the respective $J_1^{(1,2)}$ and
$\alpha^{(1,2)}$ values, are also plotted.  The lines connecting the data
points are guides to the eye.}
\label{APVOPO1kGChiMFTPars}
\end{figure}
\noindent neighbor spins in the $a$ direction by exchange constant $J_a$,
one obtains
\begin{equation}
\lambda = 2{J_a\over J_1^{\rm max}}~.
\label{EqLambda}
\end{equation}

We fitted the data in Fig.~\ref{Yama1kGChiPoly2ChnFit} by
Eqs.~(\ref{EqChiExp:all}), assuming $g = 1.969$ for all of the fits and
using Eqs.~(\ref{EqMFT:a}), (\ref{EqChiExp:c}), and~(\ref{EqMFTChi*}),
where $t\equiv \min(t^{(1)},t^{(2)})$, to determine the spin
susceptibility $\chi^{\rm spin}(T)$.  The resulting fitted parameters
$J_1^{(1,2)}$ and $\alpha^{(1,2)}$ for the chains~(1) and~(2), the rms
deviation $\sigma_{\rm rms}$ of the fit from the data, and the spin gaps
$\Delta^{(1,2)}$ for the  chains~(1) and~(2) derived using
Eqs.~(\ref{EqD(d):all}), are plotted versus $\lambda$ in
Fig.~\ref{APVOPO1kGChiMFTPars} for $-0.4 \leq \lambda \leq 0.4$ in
$\lambda$ increments of 0.05, where positive (negative) values of
$\lambda$ correspond to AF (ferromagnetic FM) coupling
$J_a$.  Not plotted in Fig.~\ref{APVOPO1kGChiMFTPars} are the fitted
values of $\chi_0$, $C_{\rm imp}$, and $\theta_{\rm imp}$, which for
$\lambda = -0.4$, 0 and~0.4 are $-$7.0(1), $-$1.99(2), 3.8(2)$\times
10^{-5}$\,cm$^3$/mol\,V, 4.65(7), 4.26(1), 3.4(1)$\times
10^{-3}$\,cm$^3$\,K/mol\,V, and $-$4.0(1), $-$3.89(2), $-$3.0(2)\,K,
respectively.  From Fig.~\ref{APVOPO1kGChiMFTPars}, a pronounced minimum
(0.095\,\%) occurs in $\sigma_{\rm rms}$ at $\lambda \approx -0.05$,
which corresponds according to Eq.~(\ref{EqLambda}) to a FM $J_a$ with
$|J_a/J_1^{(1)}| \approx 0.025$.  This is quantitatively consistent with
the above-cited estimates\cite{Garrett1997a} of this ratio based on a
one-chain model with anisotropic (in spin space) spin interactions for the
one-magnon dispersion relations observed by inelastic neutron scattering
in Ref.~\onlinecite{Garrett1997a}.  The fit to the data for $\lambda =
-0.05$ is shown as the solid curve in Fig.~\ref{AP-VOPO_MFTChiFit},
where $\chi_0 = -2.50(2)\times 10^{-5}$\,cm$^3$/mol\,V,
$C_{\rm imp} = 0.00428(1)$\,cm$^3$\,K/mol\,V,
$\theta_{\rm imp} = -3.87(2)$\,K, $J_1^{(1)}/k_{\rm B} = 139.5(2)$\,K,
$\alpha^{(1)} = 0.8597(4)$, $J_1^{(2)}/k_{\rm B} = 123.7(2)$\,K,
$\alpha^{(2)} = 0.6319(6)$, $\Delta^{(1)}/k_{\rm B} = 37.5(2)$\,K, and
$\Delta^{(2)}/k_{\rm B} = 66.1(2)$\,K\@.  In deriving the spin gap for
each chain using Eqs.~(\ref{EqD(d):all}) we have implicitly assumed that
the spin gap is unaffected by the interchain couplings.  

A somewhat more precise estimate of $\lambda$ is obtained by allowing this
parameter to vary during the fit.  The fit parameters and derived spin
gaps of the two chains~(1) and~(2) for the best fit are
\begin{mathletters}
\label{EqMFTPars:all}
\[
\chi_0 = -2.37(4)\times 10^{-5}\,{\rm cm^3\over mol\,V},~~C_{\rm imp} =
0.00427(1)\,{\rm cm^3\,K\over mol\,V},
\]
\[
\theta_{\rm imp} = -3.87(2)\,{\rm K}~,~~~\lambda = -0.037(4)~,
\]
\begin{equation}
{J_1^{(1)}\over k_{\rm B}} = 137.1(7)\,{\rm K}~,~~\alpha^{(1)} =
0.855(2)~,
\label{EqMFTPars:a}
\end{equation}
\[
{J_1^{(2)}\over k_{\rm B}} = 124.8(4)\,{\rm K}~,~~\alpha^{(2)} =
0.634(1)~,
\]
\[
{\chi^2\over {\rm DOF}} = 0.89 \Big(10^{-6}\,{\rm
cm^3\over mol\,V}\Big)^2,~~~\sigma_{\rm rms} = 0.099\,\%~,
\]
\[
{\Delta^{(1)}\over k_{\rm B}} = 37.8(5)\,{\rm K}~,~~{\Delta^{(2)}\over
k_{\rm B}} = 66.4(3)\,{\rm K}~.
\]
The spin gaps are similar
to and may be compared with those for Fit~5 in
Table~\ref{TabFitAPVOPOPowder} for the same data, in which $g = 1.969$
was also assumed but where $\lambda = 0$.  From Eq.~(\ref{EqLambda}) which
assumes a nearest neighbor interchain coordination number of~2, we obtain
the average interchain coupling strength

% Figure 6
\begin{figure}
\epsfxsize=3.4in
\centerline{\epsfbox{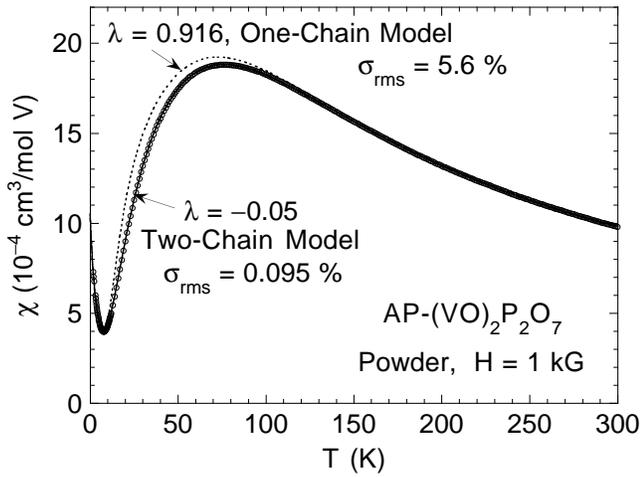}}
\vglue 0.1in
\caption{Magnetic susceptibility $\chi$ versus temperature $T$ for
a powder sample of AP-(VO)$_2$P$_2$O$_7$ ($\circ$) from
Fig~\protect\ref{Yama1kGChiPoly2ChnFit}.  The solid curve is a fit to the
data using Eq.~(\protect\ref{EqChiExp:a}) assuming a spin
susceptibility $\chi^{\rm spin}(T)$ consisting of those of equal numbers
of two $S = 1/2$ antiferromagnetic alternating-exchange Heisenberg chains
which are coupled using MFT with a ferromagnetic coupling constant
$\lambda = -0.05$.  The dotted curve is the MFT prediction using the
exchange constants found for the one-chain model by Uhrig and Normand
(Ref.~\protect\onlinecite{Uhrig1998}), for which the MFT coupling
constant is strongly antiferromagnetic with the value $\lambda = 0.916$.}
\label{AP-VOPO_MFTChiFit}
\end{figure}

\begin{equation}
{J_a\over k_{\rm B}} = {\lambda J_1^{(1)}\over 2 k_{\rm B}} \approx
-2.5\,{\rm K}~.
\label{EqMFTPars:b}
\end{equation}
\end{mathletters}

As noted in the Introduction, Uhrig and Normand\cite{Uhrig1998}
proposed a model for AP-(VO)$_2$P$_2$O$_7$ in which only one type of
alternating-exchange chain occurs and in which the (AF) interchain
couplings are given by
\begin{mathletters}
\label{EqUhrigPars:all}
\begin{equation}
{J_1\over k_{\rm B}} = 124\,{\rm K},~~\alpha = 0.793,~~{J_a\over J_1}
= 0.203,~~{J_\times\over J_1} = 0.255~,
\label{EqUhrigPars:a}
\end{equation}
where these parameters were obtained by fitting the one-magnon
inelastic neutron scattering dispersion relation data\cite{Garrett1997a}
for the lower band and using the observed Weiss
temperature\cite{Johnston1987} obtained by fitting experimental $\chi(T)$
data at high temperatures by a Curie-Weiss law (as predicted by MFT). 
Assuming that the interchain couplings do not affect the spin gap,
Eqs.~(\ref{EqD(d):all}) yield
\begin{equation}
{\Delta\over k_{\rm B}} = 44.0\,{\rm K}~.
\label{EqUhrigPars:b}
\end{equation}
This is about 16\,\% larger than our estimate for $\Delta^{(1)}$ in
Eqs.~(\ref{EqMFTPars:a}).  Of course, since their exchange constants
were determined by fitting their theory to the experimental
neutron scattering data, their spin gap is the observed
value ($\approx 36$\,K) and not that in Eq.~(\ref{EqUhrigPars:b}).  The
discrepancy arises because they find that the spin gap {\it does} depend
on the interchain couplings, as further discussed in
Sec.~\ref{SecAPVOPODispRelns} below.  

Since the interchain spin
coordination number for each of the interchain couplings is 2, the value
of the MFT interchain coupling constant predicted by
Eq.~(\ref{EqMFT:c}) is
\begin{equation}
\lambda = 2\left({J_a\over J_1} + {J_\times\over J_1}\right)~.
\label{EqUhrigPars:c}
\end{equation}
\end{mathletters}
Inserting the parameters of Uhrig and Normand in Eq.~(\ref{EqUhrigPars:a})
into this equation yields $\lambda = 0.916$.  Our fit parameters and their
variations with $\lambda$ in Fig.~\ref{APVOPO1kGChiMFTPars} argue against
this very large AF value of $\lambda$.  To further illustrate the
discrepancy within MFT between this one-chain theory and the experimental
$\chi(T)$ data, shown as the dotted curve in Fig.~\ref{AP-VOPO_MFTChiFit}
is the predicted $\chi(T)$ using $g = 1.969$, the $\chi_0$, $C_{\rm imp}$,
and $\theta_{\rm imp}$ values obtained for $\lambda = -0.05$ in
Sec.~\ref{SecAP-VOPOMFT}, and the MFT prediction
for the spin susceptibility in Eqs.~(\ref{EqMFT:all}), where
$\chi_0^*(t)$ is that of the isolated alternating-exchange chain for
which the exchange constants estimated by Uhrig and Normand in
Eq.~(\ref{EqUhrigPars:a}) were used.  The relative deviation of the
prediction from the data is \ $\sigma_{\rm rms} = 5.6$\,\%, which is about
60 times larger than the $\sigma_{\rm rms}$ obtained using the two-chain
model for $\lambda = -0.05$.  The agreement of both theoretical
predictions with the data at high temperatures is expected and in fact is
required for either model, since the MFT is most accurate at high
temperatures where it yields the Curie-Weiss law.  The significant
differences between the predictions of the two models only become
apparent at the lower temperatures.

In summary, our high-precision fits to the $\chi(T)$ data using the model
of two independent chains, in which nearest neighbor chains along the $a$
axis are coupled using MFT, indicate that the average interchain coupling
is weakly ferromagnetic, in agreement with the analysis of neutron
scattering data by Garrett {\it et al.}\cite{Garrett1997a} using a
one-chain model and in disagreement with the one-chain model of Uhrig and
Normand\cite{Uhrig1998} with strong AF interchain couplings.

\subsubsection{Single crystals}
\label{SecAPVOPOXtalChi}
\begin{figure}
\epsfxsize=3.4in
\centerline{\epsfbox{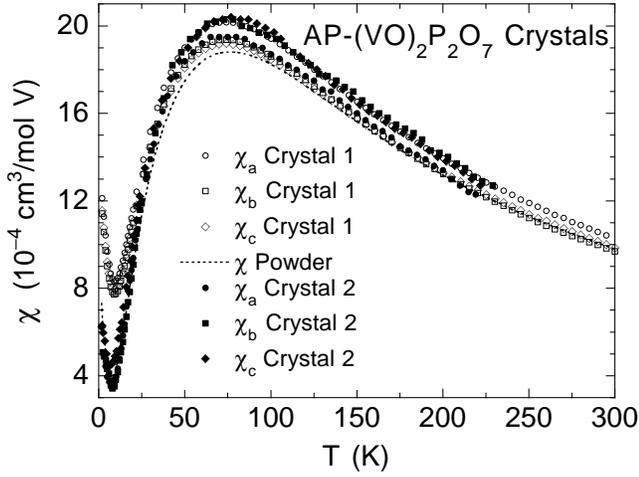}}
\vglue 0.1in
\caption{Overview of the anisotropic magnetic susceptibilities $\chi$
versus temperature
$T$ for crystals~1 (open symbols) and~2 (filled symbols, from
Ref.~\protect\onlinecite{Prokofiev1998}) of AP-(VO)$_2$P$_2$O$_7$ along
the $a$ axis (circles), $b$ axis (squares) and $c$ axis (diamonds).  Also
shown for comparison is $\chi(T)$ for the powder sample from  
Fig.~\protect\ref{Yama1kGChiPoly2ChnFit} (dotted curve).}
\label{AP-VOPOXtals}
\end{figure}

%\vglue0.18in
In this section we analyze the anisotropic $\chi(T)$ data for two single
crystals of AP-(VO)$_2$P$_2$O$_7$.  The data for a small dark green
crystal of mass 2.0(1)\,mg (``crystal~1''), measured in a magnetic field
of 5\,T using a Quantum Design SQUID magnetometer, have not been reported
previously.  There was no discernable difference between field-cooled and
zero-field-cooled $\chi(T)$ data for this crystal.  The data for crystal~2
were reported by \mbox{Prokofiev~{\it et al.}}\ in Fig.~4(a) of
Ref.~\onlinecite{Prokofiev1998} and were measured in a magnetic field of
2\,T\@.  An overview of the anisotropic $\chi(T)$ data for the two
crystals is shown in Fig.~\ref{AP-VOPOXtals}, where the
$\chi(T)$ of the powder sample in Fig.~\ref{Yama1kGChiPoly2ChnFit} is
shown for comparison as the dashed curve.  The data for the two
crystals are in agreement on a coarse scale.  The powder averages of the
data for both crystals lie above the data for the powder sample for $T
\gtrsim 25$\,K, although in the case
% Figure 7

% Figure 8
\begin{figure}
\epsfxsize=3.1in
\centerline{\epsfbox{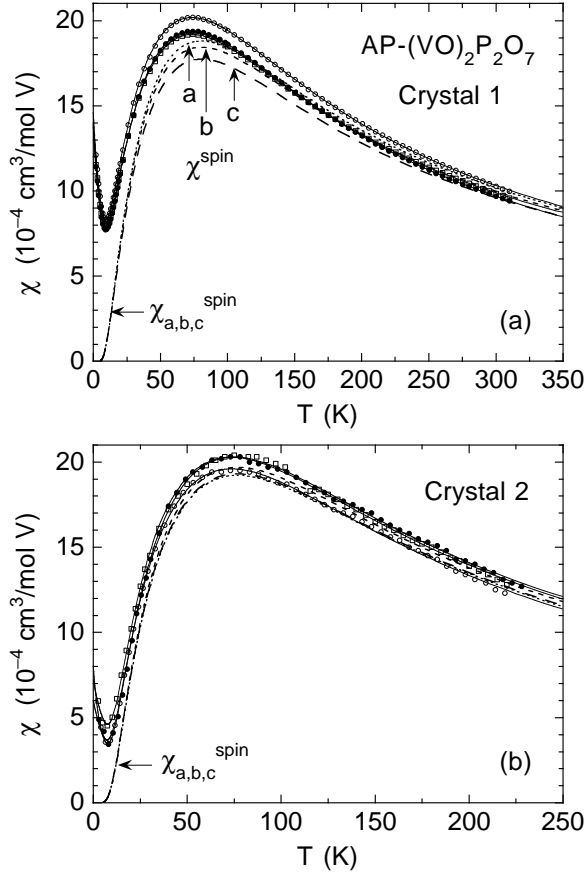}}
\vglue 0.1in
\caption{Fits to the anisotropic magnetic susceptibilities $\chi$
versus temperature $T$ for crystals~1 (a) and~2
(Ref.~\protect\onlinecite{Prokofiev1998}) (b) of AP-(VO)$_2$P$_2$O$_7$. 
The fits are shown as solid curves and the spin susceptibilities as
dashed curves.  Note that the temperature scales in~(a) and~(b) are
different.}
\label{AP-VOPOChiabcFit}
\end{figure}

\noindent  of crystal~1 this difference is not
significant since the uncertainty in the crystal mass is about 5\,\%. 
Crystal~1 shows a larger Curie-Weiss-type paramagnetic defect
and/or impurity contribution at low temperatures than crystal~2. 
Quantitative differences are seen between the anisotropic $\chi(T)$ for the two crystals.  In particular,
above about 30\,K the $\chi_a(T)$ of crystal~1 agrees with
$\chi_{b,c}(T)$ of crystal~2 and $\chi_{b,c}(T)$ of crystal~1 agrees with
$\chi_a(T)$ of crystal~2.  These qualitative anisotropy differences
cannot arise from inaccuracy in, e.g., the crystal masses, which would
only affect the respective ordinate scale.

The $a$-, $b$- and $c$-axis $\chi(T)$ data for each crystal in
Fig.~\ref{AP-VOPOXtals} were fitted simultaneously using
Eqs.~(\ref{EqChiExp:all}) by writing $\chi$ as a diagonal tensor.  We
assumed the two-chain model for the spin susceptibility of each crystal
where the values of $J_1$ and $\alpha$ for each chain are the same for all
three crystal directions, the anisotropic $g$ values are the same for
both chains, and  allowed $\chi_0$, $C_{\rm imp}$ and $\theta_{\rm imp}$
to be different for each chain and for the three field directions, for a
total of 16 fitting parameters.  The 4D fits obtained for crystals~1
and~2 are shown as the sets of three solid curves in
Figs.~\ref{AP-VOPOChiabcFit}(a) and~\ref{AP-VOPOChiabcFit}(b),
respectively, and the fitted 

\vglue-0.05in
% Table IV
\begin{table}
\caption{Fitted and derived parameters for the anisotropic $\chi(T)$ of
our crystal~1 and for crystal~2 (Ref.~\protect\onlinecite{Prokofiev1998})
of AP-(VO)$_2$P$_2$O$_7$.  A derived quantity is marked by an
asterisk~(*).  Quantities $A^{(1)}$ and $A^{(2)}$ are the quantities
associated with two independent isolated chains, respectively.}
\begin{tabular}{lcc}
Quantity & Crystal 1 & Crystal 2 \\
\hline
$\chi_{0a}\ \big({\rm 10^{-5}{cm^3\over mol\,V}}\big)$ & $-$3.4(3) &
$-$5(3) \\
$\chi_{0b}\ \big({\rm 10^{-5}{cm^3\over mol\,V}}\big)$ & $-$1.6(2) &
0(2) \\
$\chi_{0c}\ \big({\rm 10^{-5}{cm^3\over mol\,V}}\big)$ & $-$7.7(2) &
1(2) \\
$C_{{\rm imp}a}\ \big({\rm 10^{-3}\,{cm^3\,K\over mol\,V}}\big)$ &
14.9(1) & 7.1(5) \\
$C_{{\rm imp}b}\ \big({\rm 10^{-3}\,{cm^3\,K\over mol\,V}}\big)$ &
14.7(1) & 3.7(4) \\
$C_{{\rm imp}c}\ \big({\rm 10^{-3}\,{cm^3\,K\over mol\,V}}\big)$ &
13.8(1) & 7.7(5) \\
$\theta_{{\rm imp}a}$ (K) & $-$9.85(8) & $-$10.9(9) \\
$\theta_{{\rm imp}b}$ (K) & $-$10.06(8) & $-$4.5(5) \\
$\theta_{{\rm imp}c}$ (K) & $-$9.73(8) & $-$10.7(6) \\
$g_a$  & 2.003(2) & 1.98(2) \\
$g_b$  & 1.984(2) & 2.00(2) \\
$g_c$  & 1.946(2) & 1.98(2) \\
$J_1^{(1)}/k_{\rm B}$ (K) & 122(3) & 125(14) \\
$J_1^{(2)}/k_{\rm B}$ (K) & 143(4) & 128(11) \\ 
$\alpha^{(1)}$ & 0.803(9) & 0.80(5) \\
$\alpha^{(2)}$ & 0.648(6) & 0.65(4) \\
$J_2^{(1)}/k_{\rm B}$ (K) & 98(4)* & 101(16)* \\
$J_2^{(2)}/k_{\rm B}$ (K) & 93(3)* & 83(12)* \\ 
$J^{(1)}/k_{\rm B}$ (K) & 110(3)* & 113(15)* \\
$J^{(2)}/k_{\rm B}$ (K) & 118(3)* & 105(12)* \\ 
$\delta^{(1)}$ & 0.109(5)* & 0.11(3)* \\
$\delta^{(2)}$ & 0.214(5)* & 0.21(2)* \\
$\Delta^{(1)}/k_{\rm B}$ (K) & 42(3)* & 43(11)* \\
$\Delta^{(2)}/k_{\rm B}$ (K) & 74(3)* & 66(10)* \\
${\chi^2\over {\rm DOF}}$ $\big({\rm 10^{-5}{cm^3\over mol\,V}}\big)^2$
& 0.0384 & 2.0 \\
$\sigma_{\rm rms}$ (\%) & 0.159 & 1.61 \\
\end{tabular}
\label{TabFitAPVOPOXtals}
\end{table}

\noindent parameters for both crystals are given in
Table~\ref{TabFitAPVOPOXtals} where the goodnesses of fit for the two
crystals are also listed.  The anisotropic spin susceptibilities
$\chi_\alpha^{\rm spin}(T)$ were derived using Eq.~(\ref{EqChiExp:a}),
i.e., by subtracting the respective $\chi_0$ and defect and/or impurity
Curie-Weiss terms from the $\chi(T)$ fit function, and are plotted versus
temperature for crystals~1 and~2 as the two sets of three dashed curves
in Figs.~\ref{AP-VOPOChiabcFit}(a) and~\ref{AP-VOPOChiabcFit}(b),
respectively.  From Eqs.~(\ref{EqChiExp:all}), the only source of
anisotropy in $\chi^{\rm spin}(T)$ is the anisotropy in the $g$ factor.
Also listed in Table~\ref{TabFitAPVOPOXtals} are the spin gaps computed
using Eqs.~(\ref{EqD(d):all}) for the two distinct alternating-exchange
spin chains in each crystal.

Several features of the data in Table~\ref{TabFitAPVOPOXtals} are of
note.  First, as qualitatively expected from Fig.~\ref{AP-VOPOXtals}, the
concentration \ of \ paramagnetic \ defects \ and/or impurities in crystal~1 is
about a factor of two larger than in crystal~2.  Second, the spin gaps of
the two chains in each of the crystals~1 and~2 are consistent within the
error bars with each other and with those found in the above section for
the high-quality powder sample of \mbox{AP-(VO)$_2$P$_2$O$_7$} as listed
in Table~\ref{TabFitAPVOPOPowder}.  The large error bars on the fitted
parameters for crystal~2 arise in large part because the resolution in
$\chi$ for the data\cite{Prokofiev1998} above 20\,K is only
$1\times 10^{-5}$\,cm$^3$/mol\,V, which corresponds to a relative
resolution of, e.g., 1\,\% at 20\,K and 0.5\,\% at 70\,K\@.  Third, the
fitted $g_\alpha$ values are similar to, but differ in detail from, the
corresponding ESR values for \mbox{AP-(VO)$_2$P$_2$O$_7$} in
Table~\ref{TabESR}.  These discrepancies between the respective
$g_\alpha$ values may originate at least in part from the large
Curie-Weiss contribution in crystal~1 and the low resolution of the
$\chi$ data for crystal~2.

\subsection{Dispersion Relations of
AP-(VO)$_{\bbox{2}}$P$_{\bbox{2}}$O$_{\bbox{7}}$}
\label{SecAPVOPODispRelns}
\vglue0.05in

\begin{figure}
\epsfxsize=3.4in
\centerline{\epsfbox{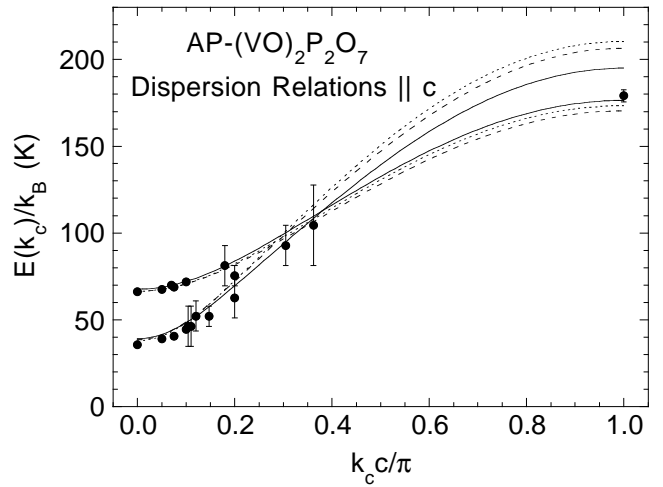}}
\vglue 0.2in
\caption{Comparison of our predicted dispersion relations of
the proposed two types of alternating-exchange chains in
AP-(VO)$_2$P$_2$O$_7$ with the dispersion relations along the
chain direction measured using inelastic neutron scattering at $T =
10$\,K by Garrett {\it et al.}\ ($\bullet$,
Ref.~\protect\onlinecite{Garrett1997a}).  The two solid curves are the
dispersion relations predicted from our exchange constants determined
from fits to $\chi(T)$ of a high-purity powder sample by the isolated
chain model, and the dashed curves are the corresponding curves for a
\mbox{MFT-coupled} chain model.  In each case, the dispersion relations
were calculated from our intrachain exchange constants using the Fourier
series to tenth order in $\alpha$ given by Knetter and
Uhrig (Ref.~\protect\onlinecite{Knetter2000}).  The dotted curve for each
chain is the dispersion relation in Eqs.~(\protect\ref{EqE(kc)Uhrig:all})
incorporating the interchain coupling according to the model of Uhrig and
Normand (Ref.~\protect\onlinecite{Uhrig1998}).}
\label{APVOPO_NaglerE(kc)Fit}
\end{figure}

The one-magnon dispersion relation $E(k_c)/J_1$ in the chain direction for
the isolated $S = 1/2$ AF alternating-exchange Heisenberg chain, with and
without a frustrating second-neighbor coupling, was recently calculated to
tenth order in $\alpha$ by Knetter and Uhrig.\cite{Knetter2000}  Thus it
is possible to make a direct and accurate comparison of the dispersion
relations predicted for the two proposed alternating-exchange chains on
the basis of our exchange constants in \mbox{AP-(VO)$_2$P$_2$O$_7$} with
those determined by Garrett {\it et al.}\cite{Garrett1997a} using
inelastic neutron scattering (INS) measurements.  For this comparison, we
first use the exchange constants for the two chains from Fit~6 in
Table~\ref{TabFitAPVOPOPowder} determined from our fit to the $\chi(T)$
data for the high-purity powder sample using the isolated chain model. 
The predicted dispersion relations for the two chains are shown as the
two solid curves in Fig.~\ref{APVOPO_NaglerE(kc)Fit} where the
experimental INS data $(\bullet)$ are also plotted.  Also shown as dashed
curves are the dispersion relations predicted for the two chains using
the intrachain exchange constants in Eq.~(\ref{EqMFTPars:a}) found from
our fit to the same $\chi(T)$ data by the same two-chain model but where
the chains are coupled along the $a$ axis using \,MFT.  \,The \,range \,of
\,our
% Figure 9
\vglue0.02in
\noindent prediction for the dispersion relation of each chain is thus 
approximately given by the region between the respective pair of solid
and dashed curves.

Our predicted dispersion relations for the two chains agree 
well with the experimental inelastic neutron scattering data in
Fig.~\ref{APVOPO_NaglerE(kc)Fit} except near the zone boundary at $k_c =
\pi/c$.  Our prediction is that two peaks in the scattered neutron
intensity versus energy at this wavevector should be seen with an energy
separation of about \mbox{20--40\,K} ($\approx$\,\mbox{2--3\,meV}),
contrary to the single data point at this wavevector in
Fig.~\ref{APVOPO_NaglerE(kc)Fit}.  However, the error bar on the
data point in Fig.~\ref{APVOPO_NaglerE(kc)Fit} at $k_c =
\pi/c$ is only the accuracy of determining the position of the
centroid of the neutron scattering peak and is not a direct measure of the
width of the peak.\cite{NaglerPC}  After most of the fits to the
$\chi(T)$ data described in this paper and the determinations of the
exchange constants were completed, we learned that recent unpublished INS
measurements on a large crystal of \mbox{AP-(VO)$_2$P$_2$O$_7$} indeed
show two peaks at this wavevector with an energy splitting of about
2\,meV,\cite{NaglerPC} which partially confirms our prediction.  We also
predict from Fig.~\ref{APVOPO_NaglerE(kc)Fit} that the dispersion
relations of the two chains should cross within the intermediate
wavevector regime.  We are not aware of experimental INS data that
address this aspect of the dispersion relations.

% Figure 10
\begin{figure}
\epsfxsize=3.4in
\centerline{\epsfbox{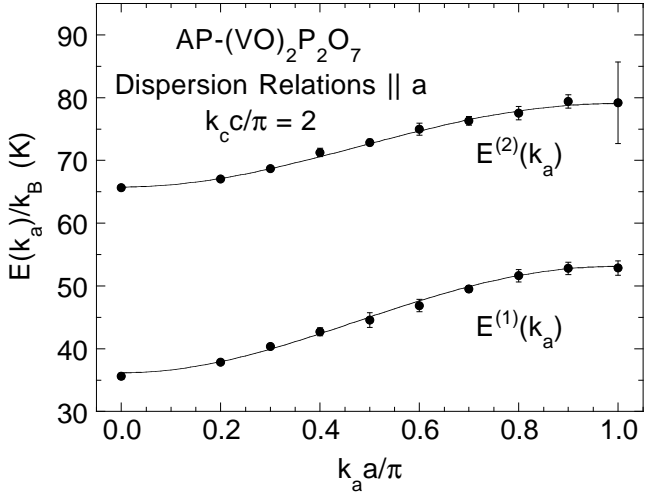}}
\vglue 0.2in
\caption{Fits to the dispersion relations of
the proposed two types of alternating-exchange chains in
AP-(VO)$_2$P$_2$O$_7$ in the $a$ direction and perpendicular to the
chain direction, measured using inelastic neutron scattering at $T =
10$\,K by Garrett {\it et al.}\ ($\bullet$,
Ref.~\protect\onlinecite{Garrett1997a}).  The two solid curves are fits to
the respective data by Eqs.~(\protect\ref{EqE(k)Uhrig:all}), which is the 
special case for $k_c = 0$ of the general dispersion relation calculated
by Uhrig and Normand (Ref.~\protect\onlinecite{Uhrig1998}).}
\label{APVOPO_NaglerE(ka)Fits}
\end{figure}
\vglue0.06in
%\noindent 
We note that the two sets of exchange constants in
Table~\ref{TabFitAPVOPOXtals} for the two chains in crystal~1 also
predict an energy splitting of the neutron peak at $k_c = \pi/c$, but in
this case the predicted dispersion relations of the two chains do not
cross.  We consider the experimental data and modeling for the
high-purity powder sample to be more reliable than for this crystal,
for reasons discussed in Sec.~\ref{SecAPVOPOXtalChi} above, and hence
expect that the dispersion relations of the two chains will ultimately be
observed to cross. 

A quantitative estimate of the interchain couplings $J_a$ and
$J_\times$ can be obtained by fitting the observed\cite{Garrett1997a}
dispersion relation for each of the two chains in the $a$ direction,
shown in Fig.~\ref{APVOPO_NaglerE(ka)Fits}, by the prediction of Uhrig and
Normand\cite{Uhrig1998}
\begin{mathletters}
\label{EqE(k)Uhrig:all}
\begin{eqnarray}
E(k_a,k_c&=&0) = \Delta(J_1,\alpha,\mu_+,\mu_-) \nonumber\\
&+& J_1\sum_{n=1}^\infty a_n(\alpha,\mu_+,\mu_-)[\cos(n k_aa) - 1]
\label{EqE(k)Uhrig:a}
\end{eqnarray}
where $\mu_\pm \equiv (J_a \pm J_\times)/J_1$.  To third order in
$\alpha$, $\mu_-$, and/or $\mu_+$ their general dispersion relation yields
\[
a_1 = {\mu_-\over 4}\Big[4 + \alpha(2 + \mu_+) - \alpha^2 - \mu_-^2\Big]~,
\]
\begin{equation}
a_2 = -{\mu_-^2\over 8}(2 + 3\alpha + 2\mu_+)~,~~~~a_3 =
{\mu_-^3\over 8}~.
\label{EqE(k)Uhrig:b}
\end{equation}
\end{mathletters}

We found that the data in Fig.~\ref{APVOPO_NaglerE(ka)Fits} could not be
fitted by Eqs.~(\ref{EqE(k)Uhrig:all}) assuming $J_\times = 0$ and
$J_a/J_1 \approx -0.019$ as inferred from $\lambda \approx -0.037$ in
Eqs.~(\ref{EqMFTPars:all}).  However, we can still retain this 
experimentally determined value of $\lambda$ by allowing $J_\times$ to be
nonzero.  In particular, from Eq.~(\ref{EqUhrigPars:c}) one obtains
$\mu_+ = \lambda/2$.  Therefore we set $\mu_+ = -0.019$ and used the
experimentally determined values of $J_1$ and $\alpha$ for each chain in
Eq.~(\ref{EqMFTPars:a}) in the fit to the respective dispersion
relation.  This left $\Delta$ and $\mu_-$ as the only adjustable
parameters.  We did not use the experimentally determined $\Delta$ values
in Eqs.~(\ref{EqMFTPars:all}) because the total width of the
experimental dispersion relation for each chain is relatively small and
the small difference between the experimental
$\Delta$ value and the neutron scattering result would give a
significant systematic shift to the fit (see also below).  The fits are
shown as the solid curves in Fig.~\ref{APVOPO_NaglerE(ka)Fits} for
chains~(1) and~(2), respectively, for which the parameters are
\[
{J_a^{(1)}\over J_1} = -0.035~,~~~{J_\times^{(1)}\over J_1} =
+0.016~,~~~{\Delta^{(1)}\over k_{\rm B}} = 36.2\,{\rm K}~,
\]
\begin{equation}
{J_a^{(2)}\over J_1} = -0.032~,~~~{J_\times^{(2)}\over J_1} =
+0.013~,~~~{\Delta^{(2)}\over k_{\rm B}} = 65.7\,{\rm K}~.
\label{EqE(ka)Pars}
\end{equation}
Thus we find that a small but finite AF value of the interchain
interaction $J_\times$ is necessary to fit the dispersion data
perpendicular to the alternating-exchange chains if we retain the fitted
$\lambda$ value in Eqs.~(\ref{EqMFTPars:all}).  Since the
nearest-neighbor interchain interaction $J_a$ is ferromagnetic, the
next-nearest-neighbor AF interchain interaction 
$J_\times$ is not a geometrically frustrating interaction.  

On the other hand, an equally good and nearly identical fit for each chain
as shown in Fig.~\ref{APVOPO_NaglerE(ka)Fits} can be obtained assuming
that $J_\times = 0$ if we relax the above condition on $\lambda$.  In
this case we still use the exchange constants in
Eqs.~(\ref{EqMFTPars:all}) but we set $\mu_- = \mu_+ \equiv \mu$ in the
fit to the transverse dispersion data for each chain, yielding the same
respective gap values as in Eqs.~(\ref{EqE(ka)Pars}), but where 
$\mu^{(1)} = -0.050(2)$ and $\mu^{(2)} = -0.044(2)$, so that the
$J_a/J_1 = \mu$ and  $\lambda = 2\mu \approx -0.10$ values are larger in
magnitude than in Eqs.~(\ref{EqE(ka)Pars}).

The dispersion relation parallel to a chain ($||\,c$) calculated by Uhrig
and Normand,\cite{Uhrig1998} in which the $J_a$ and $J_\times$ interchain
couplings are included to third order in $\alpha$, $\mu_-$, and/or
$\mu_+$, yields
\begin{mathletters}
\label{EqE(kc)Uhrig:all}
\begin{eqnarray}
E(k_a&=&0,k_c) = \Delta + [E_0(k_a=0,k_c) - E_0(0,0)] \nonumber\\
&+& J_1\sum_{n=1}^\infty b_n(\alpha,\mu_+,\mu_-)[\cos(n k_cc) - 1]
\label{EqE(kc)Uhrig:a}
\end{eqnarray}
with
\begin{equation}
b_1 = {\mu_-\alpha\over 2}\Big[1 - {\alpha\over 4} + {\mu_+ -
\mu_-\over 2}\Big],~~b_2 = {3\mu_-\alpha^2\over 16}~,
\label{EqE(kc)Uhrig:b}
\end{equation}
\end{mathletters}
where $E_0(0,k_c)$ is the dispersion relation for the isolated 
alternating-exchange chain to tenth order in
$\alpha$.\cite{Knetter2000}  The resulting dispersion relations calculated
for the two chains using Eqs.~(\ref{EqE(kc)Uhrig:all}), with the the
intrachain exchange constants and spin gaps in Eqs.~(\ref{EqMFTPars:all})
and the interchain couplings given in Eqs.~(\ref{EqE(ka)Pars}), are shown
as the dotted curves in Fig.~\ref{APVOPO_NaglerE(kc)Fit}.  These are
respectively very similar to those for the MFT-coupled chain parameters
already plotted as the dashed curves in Fig.~\ref{APVOPO_NaglerE(kc)Fit}.

An inconsistency in our fit to the transverse dispersion relation for each
chain is that the calculated spin gap is smaller than the observed and
fitted one in Fig.~\ref{APVOPO_NaglerE(ka)Fits}.  The spin gap in
Eq.~(\ref{EqE(k)Uhrig:a}) is given by the third order expansion of Uhrig
and Normand\cite{Uhrig1998} as
\begin{eqnarray}
\Delta = J_1\Bigg\{{\Delta_0(\alpha)\over J_1} &+& {\mu_-\over 4}\Big[4 +
\alpha(2 + \mu_+) - \alpha^2\Big] \nonumber\\ 
&+& {\mu_-^2\over 8}(4 - 2\alpha + \mu_+) - {\mu_-^3\over 8}\Bigg\}~,
\label{EqGapUhrig}
\end{eqnarray}
where $\Delta_0(\alpha)/J_1$ is the spin gap in
Eqs.~(\ref{EqDimParams2:all}) or~(\ref{EqD(d):all}) in the absence of
interchain coupling.  For the exchange constants in
Eqs.~(\ref{EqMFTPars:a}) and (\ref{EqE(ka)Pars}) used to fit the
transverse dispersion relations in Fig.~\ref{APVOPO_NaglerE(ka)Fits} for
chains~(1) and~(2), from Eq.~(\ref{EqGapUhrig}) we obtain
$\Delta^{(1)}/k_{\rm B} = 29.2$\,K and 
$\Delta^{(2)}/k_{\rm B} = 59.8$\,K, respectively.  These spin gaps are
each significantly smaller than the fitted ones in
Eqs.~(\ref{EqE(ka)Pars}), respectively.  These discrepancies arise
because the interchain couplings change the spin gap, contrary to our
implicit assumption when we fitted the experimental $\chi(T)$ data using
the \mbox{MFT-coupled} chain model, so the $J_1$ and $\alpha$ intrachain
exchange parameters for each chain derived from the MFT fit to these data
must be considered in the present model to be effective values.  The
degree to which the effective exchange constants differ from the actual
ones is difficult to evaluate.  The combined analysis we have done of the
susceptibility and dispersion relations is as rigorous as can be done
without having in hand an accurate theoretical expression for the spin
susceptibility of the system which includes the influence of interchain
couplings and concommitant changes in the two spin gaps.

\subsection{Magnetic Susceptibility of
HP-(VO)$_{\bbox{2}}$P$_{\bbox{2}}$O$_{\bbox{7}}$}
\label{SecChiHPVOPO}

\subsubsection{Crushed crystals}
\label{SecCrushedXtalChi}

\begin{figure}
\epsfxsize=3.4in
\centerline{\epsfbox{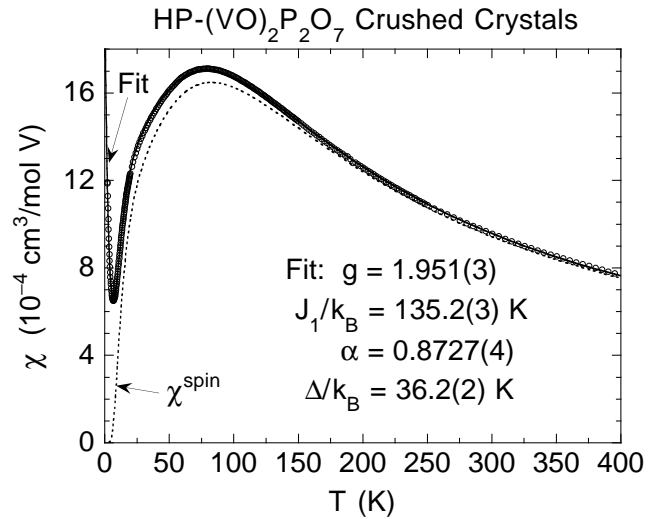}}
\vglue 0.1in
\caption{Magnetic susceptibility $\chi$ versus temperature $T$ for
a sample of HP-(VO)$_2$P$_2$O$_7$ crushed single crystals ($\circ$) from
2 to 400\,K\@.  The solid curve is a two-dimensional fit to
the 300 data points using Eq.~(\protect\ref{EqChiExp:a}) and the $S = 1/2$
antiferromagnetic alternating-exchange Heisenberg chain model for the
intrinsic spin susceptibility $\chi^{\rm spin}(T)$.  The
dotted curve is the fitted $\chi^{\rm spin}(T)$.  The fitted values of the
$g$ factor, the larger of the two exchange constants in the
alternating-exchange chain $J_1$, the alternation parameter $\alpha\equiv
J_2/J_1$, and the derived spin gap $\Delta$ are listed.}
\label{HP-VOPOCrushedXtalsChiFit}
\end{figure}

We begin our modeling of $\chi(T)$ for HP-(VO)$_2$P$_2$O$_7$ using data
shown in Fig.~\ref{HP-VOPOCrushedXtalsChiFit} which we obtained for a
72.2\,mg sample of crushed green transparent single crystals.  These data
are expected to be more accurate and yield more reliable values of the
exchange constants and spin gap than the fits to the data for a powder
and for a very small single crystal discussed in the following two
sections, respectively.  The data were modeled using
Eqs.~(\ref{EqChiExp:all}) in which $\chi^*(t)$ is the theoretical reduced
susceptibility for the $S = 1/2$ AF alternating-exchange Heisenberg chain
as proposed by Azuma {\it et al.}\cite{Azuma1999}  \ The fit  
% Figure 11
\vglue0.32in
% Table V
\begin{table}
\caption{Fitted and derived parameters for $\chi(T)$ of
crushed single crystals of HP-(VO)$_2$P$_2$O$_7$ obtained using the
one-chain and two-chain models.  A derived quantity is marked by an
asterisk (*).  Quantities $A^{(1)}$ and $A^{(2)}$ are the quantities
associated with two independent isolated chains, respectively.}
\begin{tabular}{lcc}
Quantity & One-Chain Model & Two-Chain Model \\
\hline
$\chi_{0}\ \big({\rm 10^{-5}{cm^3\over mol\,V}}\big)$ & $-$0.1(4) &
$-$1.0(3) \\
$C_{\rm imp}\ \big({\rm 10^{-3}\,{cm^3\,K\over mol\,V}}\big)$ &
4.88(6) & 4.70(4) \\
$\theta_{\rm imp}$ (K) & $-$1.99(5) & $-$1.79(3) \\
$g$  & 1.951(3) & 1.958(3) \\
$J_1^{(1)}/k_{\rm B}$ (K) & 135.2(3) & 134(14) \\
$J_1^{(2)}/k_{\rm B}$ (K) & --- & 136(14) \\ 
$\alpha^{(1)}$ & 0.8727(4) & 0.90(2) \\
$\alpha^{(2)}$ & --- & 0.85(2) \\
$J_2^{(1)}/k_{\rm B}$ (K) & 118.0(3)* & 120(15)* \\
$J_2^{(2)}/k_{\rm B}$ (K) & --- & 115(15)* \\ 
$J^{(1)}/k_{\rm B}$ (K) & 126.6(3)* & 127(15)* \\
$J^{(2)}/k_{\rm B}$ (K) & --- & 126(14)* \\ 
$\delta^{(1)}$ & 0.0680(3)* & 0.054(10)* \\
$\delta^{(2)}$ & --- & 0.082(10)* \\
$\Delta^{(1)}/k_{\rm B}$ (K) & 33.9(2)* & 29(7)* \\
$\Delta^{(2)}/k_{\rm B}$ (K) & --- & 39(7)* \\
${\chi^2\over {\rm DOF}}$ $\big({\rm 10^{-5}{cm^3\over mol\,V}}\big)^2$
& 0.36 & 0.16 \\
$\sigma_{\rm rms}$ (\%) & 0.58 & 0.35 \\
\end{tabular}
\label{TabFitHPVOPOPowdXtal}
\end{table}
\noindent is shown as
the solid curve in Fig.~\ref{HP-VOPOCrushedXtalsChiFit} and the fitted
$\chi^{\rm spin}(T)$ is shown as the dotted curve.  The fitted $g$,
$J_1$, and $\alpha$ values are listed in the figure, along with the spin
gap computed using Eqs.~(\ref{EqD(d):all}).  The other parameters of the
fit are shown in Table~\ref{TabFitHPVOPOPowdXtal}.  The fitted $g$ value
is very close to the powder average value 1.958 in Table~\ref{TabESR}
obtained from ESR measurements.  The impurity Curie constant is
equivalent to the contribution of 1.3\,mol\,\% with respect to V of
spins~1/2 with $g = 2$.

We used the data set in Fig.~\ref{HP-VOPOCrushedXtalsChiFit} to estimate 
typical nonstatistical errors that may arise when using the two-chain
model to fit the $\chi(T)$ data for AP-(VO)$_2$P$_2$O$_7$ in the above two
sections.  In addition, if the present one-chain model for
HP-(VO)$_2$P$_2$O$_7$ is appropriate, then a fit by the two-chain model
should yield very similar exchange constants and spin gaps for the two
chains of the model, which ideally would be respectively identical for
the two chains.  The parameters of the two-chain fit are compared with
those of the above single-chain fit in
\mbox{Table}~\ref{TabFitHPVOPOPowdXtal}.  We see that the fitted
parameters of the two chains using the two-chain model are the same
within the limits of error with each other and with the parameters of the
single-chain model, respectively.  This result indirectly confirms that
the large differences between the exchange constants and spin gaps found
above for the two chains in AP-(VO)$_2$P$_2$O$_7$ are reliable.

\subsubsection{MFT analysis of interchain coupling}
\label{SecHPVOPOMFT}

\begin{figure}
\epsfxsize=3.4in
\centerline{\epsfbox{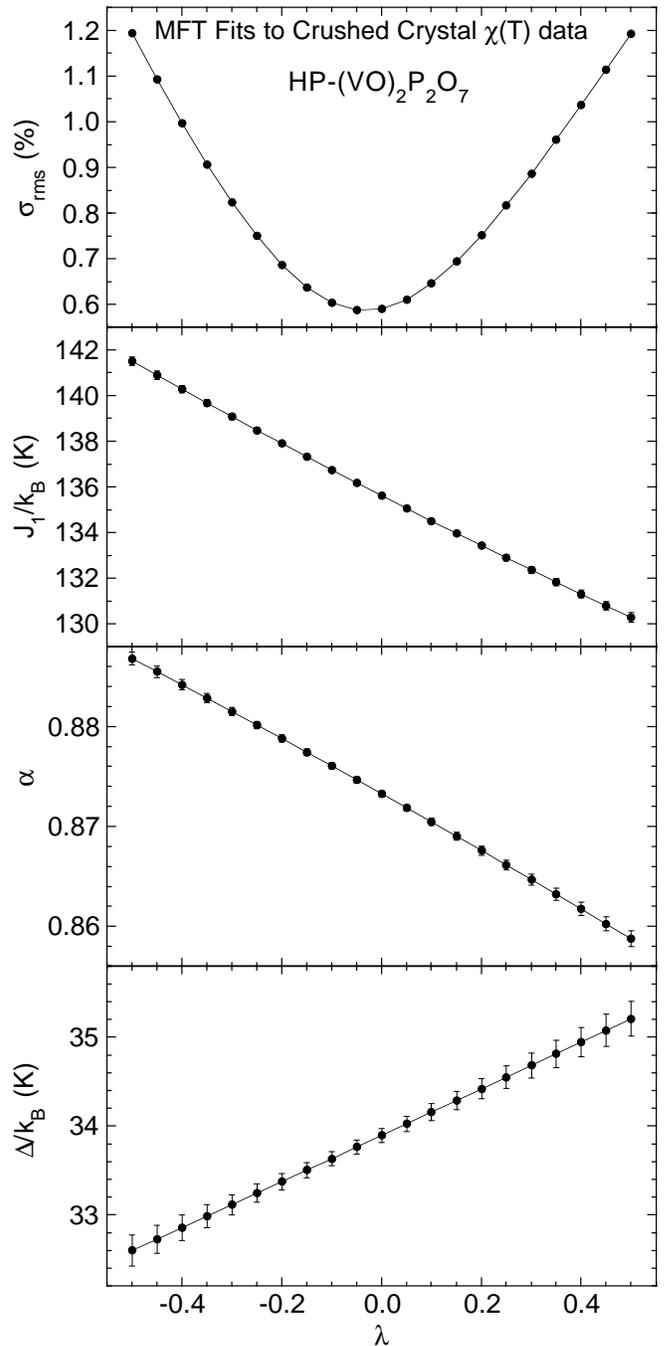}}
\vglue 0.53in
\caption{Parameters $J_1$ and $\alpha$ and the rms deviation
$\sigma_{\rm rms}$ of the fits to the $\chi(T)$ data for crushed
crystals of HP-(VO)$_2$P$_2$O$_7$ in
Fig.~\protect\ref{HP-VOPOCrushedXtalsChiFit} by MFT for coupled
alternating-exchange chains versus the MFT
interchain coupling constant $\lambda$.  A $g = 1.958$ was assumed in all
of the fits.  The spin gap $\Delta$, derived from the 
$J_1$ and $\alpha$ values for each $\lambda$, is also plotted versus
$\lambda$.  The error bars are shown for $J_1$, $\alpha$, and $\Delta$. 
The lines connecting the data points are guides to the eye.}
\label{HP-VOPOPowdXtalMFTPars}
\end{figure}

There are no data available for the strength of the interchain
coupling in HP-(VO)$_2$P$_2$O$_7$.  We estimate this
coupling in the same way as in Sec.~\ref{SecAP-VOPOMFT} for
\mbox{AP-(VO)$_2$P$_2$O$_7$}.  The most precise and accurate $\chi(T)$
data available for HP-(VO)$_2$P$_2$O$_7$ are those for the crushed crystal
sample presented and discussed in the previous section.  We determined the
fitting parameters $\chi_0$, $C_{\rm imp}$, $\theta_{\rm imp}$, $J_1$, and
$\alpha$ as a function of the MFT interchain coupling constant $\lambda$
over the range $-0.5 \leq \lambda \leq 0.5$, where the fixed $g = 1.958$
was assumed in all these one-chain model fits.  The $\sigma_{\rm rms}$ and
the $J_1$ and $\alpha$ parameters are plotted versus $\lambda$ in
Fig.~\ref{HP-VOPOPowdXtalMFTPars}, along with the spin gap $\Delta$ determined from $J_1$ and $\alpha$ using
Eqs.~(\ref{EqD(d):all}).  At
$\lambda = -$0.5, 0, and~0.5, the fitted parameters
$\chi_0$, $C_{\rm imp}$, and $\theta_{\rm imp}$ were, respectively,
$-$6.1(2), $-$0.86(9), 3.7(2)$\times 10^{-5}$\,cm$^3$/mol\,V, 5.52(9),
4.96(4), 4.53(8)$\times 10^{-3}$\,cm$^3$\,K/mol\,V, and $-$2.33(9),
$-$2.03(4), $-$1.81(8)\,K\@.  From the top panel of
Fig.~\ref{HP-VOPOPowdXtalMFTPars}, the $\sigma_{\rm rms}$ shows 
an approximately parabolic variation with $\lambda$, with a minimum at
$\lambda \approx -0.05$, indicating  a weak ferromagnetic interchain
coupling as also deduced above for AP-(VO)$_2$P$_2$O$_7$.  

We next allowed $\lambda$ to vary during the fit to determine a more
precise value.  The fit parameters and derived spin gap of the
alternating-exchange chain for the best fit are
% Figure 12
%\vglue0.12in

\newpage
\begin{mathletters}
\label{EqHPMFTPars:all}
\[
\chi_0 = -1.4(2)\times 10^{-5}\,{\rm cm^3\over mol\,V},~~C_{\rm imp} =
0.00502(4)\,{\rm cm^3\,K\over mol\,V},
\]
\[
\theta_{\rm imp} = -2.06(4)\,{\rm K}~,~~~\lambda = -0.055(14)~,
\]
\begin{equation}
{J_1\over k_{\rm B}} = 136.2(3)\,{\rm K}~,~~\alpha =
0.8748(5)~,
\label{EqHPMFTPars:a}
\end{equation}
\[
{\chi^2\over {\rm DOF}} = 0.34 \Big(10^{-5}\,{\rm
cm^3\over mol\,V}\Big)^2,~~~\sigma_{\rm rms} = 0.59\,\%~,
\]
\[
{\Delta\over k_{\rm B}} = 33.8(2)\,{\rm K}~.
\]
The spin gap is identical with that obtained for the one-chain fit in
Table~\ref{TabFitHPVOPOPowdXtal} for the same data, in which $g = 1.958$
was also assumed but where $\lambda = 0$, and the other parameters are
also very similar, respectively.  From Eq.~(\ref{EqLambda}), we obtain the
average interchain coupling strength
\begin{equation}
J_a = {\lambda J_1\over 2 k_{\rm B}} \approx -3.7\,{\rm K}~.
\end{equation}
\end{mathletters}

\subsubsection{Powder sample: low-$T$ fits}
\label{SecPowderChi}
\begin{figure}
\epsfxsize=3.4in
\centerline{\epsfbox{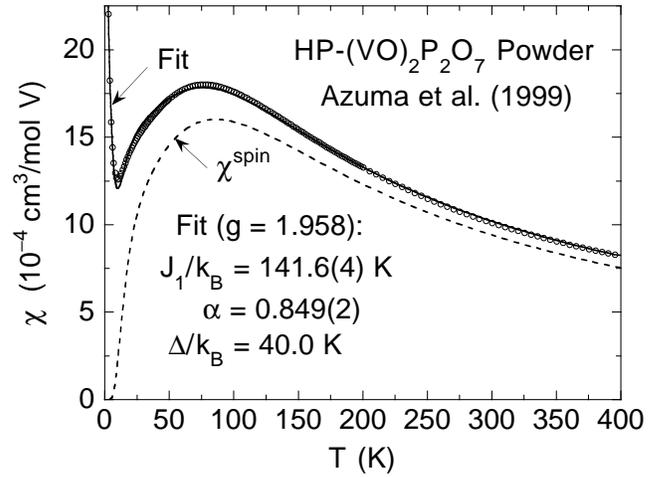}}
\vglue 0.1in
\caption{Magnetic susceptibility $\chi$ versus temperature $T$ for
a powder sample of HP-(VO)$_2$P$_2$O$_7$ ($\circ$) from 2\,K to
400\,K (Ref.~\protect\onlinecite{Azuma1999}).  The solid curve is a
two-dimensional fit to the 164 data points using
Eq.~(\protect\ref{EqChiExp:a}) and the $S = 1/2$ antiferromagnetic
alternating-exchange Heisenberg chain model for the intrinsic spin
susceptibility $\chi^{\rm spin}(T)$, using the powder-averaged $g$-value
of 1.958 determined from ESR measurements.  The dashed curve is the fitted
$\chi^{\rm spin}(T)$.  The fitted values of the larger of the two
exchange constants in the alternating-exchange chain $J_1$, the
alternation parameter $\alpha\equiv J_2/J_1$ and the derived spin gap
$\Delta$ are listed.}
\label{HP-VOPOChiFit}
\end{figure}

\vglue-0.01in
The $\chi(T)$ of a powder sample of HP-(VO)$_2$P$_2$O$_7$ was previously
reported by \mbox{Azuma {\it et al.}},\cite{Azuma1999} shown as the open
circles in Fig.~\ref{HP-VOPOChiFit}.  A fit of the data up to 30\,K by
Eqs.~(\ref{EqChiExp:all}), where $\chi^*(t)$ is the low-$T$
approximation for the spin susceptibility of a gapped 1D $S = 1/2$ spin
system in Eq.~(\ref{EqTroyer:a}), yielded the spin gap $\Delta/k_{\rm B} =
27$\,K\@.\cite{Azuma1999}  In this fit, the parameter $A$ in
Eq.~(\ref{EqTroyer:a}) was treated as an independently adjustable
parameter.  In this section we carry out a precise fit of the same data
set by Eqs.~(\ref{EqChiExp:all}) using the accurately known
$\chi^*(t,\alpha)$ spin susceptibility prediction for the $S = 1/2$ AF
alternating-exchange Heisenberg chain.\cite{Johnston1999}  The present
fit was carried out in order to compare the fitted parameters
respectively obtained from the two types of fits to the same $\chi(T)$
data set for the same sample.

We fitted the $\chi(T)$ data in Fig.~\ref{HP-VOPOChiFit} using
Eqs.~(\ref{EqChiExp:all}), where $\chi^*(t,\alpha)$ is that for the $S =
1/2$ AF alternating-exchange Heisenberg chain given in
Ref.~\onlinecite{Johnston1999}, and with the $g$ value fixed at the
spherically-averaged value 1.958 determined from single-crystal
anisotropic ESR measurements (see Table~\ref{TabESR}).  The resulting fit
is shown as the solid curve in Fig.~\ref{HP-VOPOChiFit}, and the fitted
spin susceptibility is shown as the dashed curve.  The parameters of the
fit are
\[
\chi_0 = 3.8(3) \times 10^{-5}\,{\rm {cm^3\over mol\,V}}~,
\]
\[ C_{\rm imp} = 0.0121(2)\,{\rm {cm^3\,K\over mol\,V}}~,~~\theta =
-2.6(1)\,{\rm K},
\]
% Figure 13
%\noindent

\begin{equation} {J_1\over k_{\rm B}} = 141.6(4)\,{\rm
K}~,~~\alpha = 0.849(2)~.
\label{EqChiPowdPars}
\end{equation}

The $C_{\rm imp}$ value in Eq.~(\ref{EqChiPowdPars}) is rather large,
equivalent to the contribution of 3.2\,mol\% of spins-1/2 with respect to
V and with $g = 2$.  Therefore, the fitted prefactor $1/J_1$ to
$\chi^*(t)$ in Eq.~(\ref{EqChiExp:c}) could be too small by about 3\% if
the magnetic species in the impurity phase is $S = 1/2$ V$^{+4}$.  On the
other hand, if the impurities/defects have a spin larger than 1/2, the
influence on the fitted parameters could be much smaller.  In order to
test this possible influence, we next allowed the $g$ value to be an
adjustable parameter in the fit, which has the effect of allowing the
amount of V in the HP-(VO)$_2$P$_2$O$_7$ phase relative to that in the
impurity phase to be variable.  The parameters obtained,

\[
\chi_0 = 0.0(12) \times 10^{-5}\,{\rm {cm^3\over mol\,V}}~,
\]
\[ C_{\rm imp} = 0.0125(3)\,{\rm {cm^3\,K\over mol\,V}}~,~~\theta =
-2.7(1)\,{\rm K},
\]
\begin{equation} g = 2.00(1)~,~~~{J_1\over k_{\rm B}} = 143.9(8)\,{\rm
K}~,~~\alpha = 0.854(2)~,
\label{EqChiPowdPars2}
\end{equation}
are very similar to those obtained in the above fit with fixed $g$.  The
fitted value of the prefactor $g^2/(J_1/k_{\rm B})$ to $\chi^*(t)$ in
Eq.~(\ref{EqChiExp:c}) increased, as anticipated, from 2.71\,K$^{-1}$ to
2.77\,K$^{-1}$.  Although the fitted $g$ value increased slightly from
the value used in the first fit, it \,is 
% Figure 14
\begin{figure}
\epsfxsize=3.4in
\centerline{\epsfbox{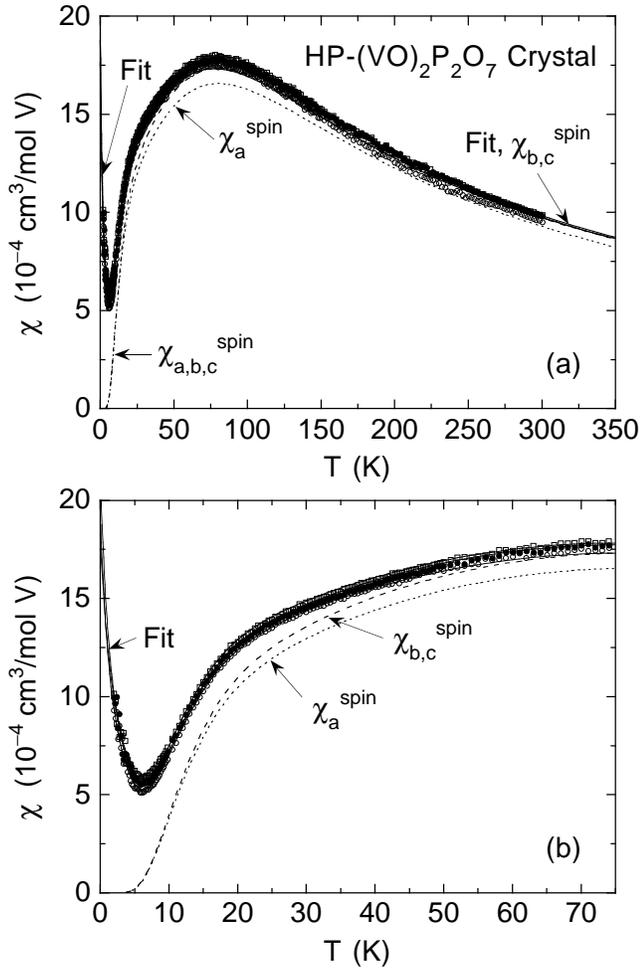}}
\vglue 0.1in
\caption{(a) Magnetic susceptibility $\chi$ versus temperature $T$ for
single crystal HP-(VO)$_2$P$_2$O$_7$ along the $a$-axis ($\circ$),
$b$-axis ($\bullet$) and $c$-axis (open squares) directions from
2\,K to 300\,K\@.  \mbox{(b)  Expanded} plot of the data in (a) at low
temperatures below the broad maximum in $\chi(T)$ at about 75\,K\@.  In
(a) and~(b), the set of three solid curves is a four-dimensional fit to
all 885 data points for the $a$-, $b$- and $c$-axis directions using
Eq.~(\protect\ref{EqChiExp:a}) and the $S = 1/2$ antiferromagnetic
alternating-exchange Heisenberg chain model for the intrinsic spin
susceptibility $\chi^{\rm spin}(T)$.  The dashed curves are the fitted
$\chi_\alpha^{\rm spin}(T)$ for the magnetic field along the $\alpha =
a$-axis (short dash) and $b$- and $c$-axes (longer dash); the fitted
$\chi_b^{\rm spin}(T)$ and $\chi_c^{\rm spin}(T)$ are indistinguishable
on the scale of the figure.}
\label{HPVOPOXtalChiabc}
\end{figure}
\vglue0.04in
\noindent  
still close to 2.  We conclude
that the magnetic impurities and/or defects giving rise to the
Curie-Weiss term have little influence on the fitted exchange constants
in the HP-(VO)$_2$P$_2$O$_7$ phase in the sample.

Taking $J_1/k_{\rm B} = 142$\,K and $\alpha = 0.85$ from
Eq.~(\ref{EqChiPowdPars}), Eq.~(\ref{EqDimParams2:a}) yields the spin gap
$\Delta/k_{\rm B} = 40$\,K, about 50\% larger than the above value of
27\,K estimated by \mbox{Azuma {\it et al.}}\cite{Azuma1999} by fitting
the same data up 30\,K (i.e., up to a $T\sim\Delta/k_{\rm B}$) using the
low-$T$ approximation to the spin susceptibility
in Eq.~(\ref{EqTroyer:a}).  Thus from the respective values of $\Delta$,
we find that the error arising from estimating the gap value by fitting
low-$T$ $\chi(T)$ data using the low-$T$ approximation for the spin
susceptibility is about 50\% in this case.  The temperature range over
which the low-$T$ approximation is fitted to the experimental data is
expected to influence this error (see Sec.~\ref{SecChiPwdAve}).

\subsubsection{Single crystal}
\label{SecXtalChi}
%\vglue0.12in
The $\chi(T)$ data reported by Saito {\it et al.}\cite{Saito2000} for a
0.26\,mg single crystal of HP-(VO)$_2$P$_2$O$_7$ are plotted in
Fig.~\ref{HPVOPOXtalChiabc}.  By comparison of these data with those for
the powder sample in Fig.~\ref{HP-VOPOChiFit}, the Curie-Weiss impurity
and/or defect contribution to $\chi(T)$ for the crystal at low
temperatures is seen to be significantly smaller than for the powder
sample.  From our fit below, we find that it is in fact about a factor of
three smaller.  This much smaller impurity contribution enhances the
reliability of the fitted exchange constants and the derived spin gap
obtained from modeling the data for the single crystal.  At higher
temperatures, the magnitude of the powder-averaged $\chi(T)$ for the
single crystal is very similar to  the $\chi(T)$ for the powder, as would
have been expected.

The modeling of the single-crystal $\chi(T)$ data was carried out in
a similar way as for the two crystals of \mbox{AP-(VO)$_2$P$_2$O$_7$} in
Sec.~\ref{SecAPVOPOXtalChi}, except that here we use a one-chain model
instead of a two-chain model for the spin susceptibility.  All 885 data
points of the $a$-, $b$-, and $c$-axis $\chi(T)$ data sets in
Fig.~\ref{HPVOPOXtalChiabc} were fitted simultaneously using
Eqs.~(\ref{EqChiExp:all}) by writing $\chi$ as a diagonal tensor and using
the three fixed $g$ values determined for fields along the three
principal axis directions from ESR measurements\cite{Saito2000} as given
in \mbox{Table}~\ref{TabESR} above, respectively.  With eleven fitting
parameters, the data to  parameter ratio is~80.  The four-dimensional fit
obtained is shown as the set of three solid curves in
Fig.~\ref{HPVOPOXtalChiabc}, and the fitted parameters are given in

%\vglue0.1in
% Table VI
\begin{table}
\caption{Fitted parameters for the anisotropic magnetic susceptibility
of single crystal HP-(VO)$_2$P$_2$O$_7$ along the $a$-, $b$-, and
$c$-axis directions from 2 to 300\,K using the one-chain model for
the spin susceptibility.  Derived quantities are shown with an asterisk
(*). The values of $J_1$, $J_2$, $\alpha$, $J$, $\delta$, and $\Delta$ are
the same for all three crystal axis directions.}
\begin{tabular}{lccc} 
Quantity & $a$-axis & $b$-axis & $c$-axis \\
\hline
$\chi_{0} \big(10^{-5}\,{\rm{cm^3\over mol\,V}}\big)$ & 3.4(2) &
$-$0.9(2) & 0.0(2) \\
$C_{\rm imp} \big(10^{-3}\,{\rm{cm^3\,K\over mol\,V}}\big)$ & 3.40(6) &
3.90(6) & 4.39(7) \\
$\theta_{\rm imp}$\,(K) & $-$2.48(9) & $-$1.98(8) & $-$2.75(9) \\
$J_1/k_{\rm B}$\,(K) & & 131.6(1) \\
$\alpha$ & & 0.8709(5) \\
$\chi^{\rm VV} \big(10^{-5}\,{\rm{cm^3\over mol\,V}}\big)$ & 9.5* & 5.2* &
6.1* \\
$J_2/k_{\rm B}$\,(K) & & 114.6(2)* \\
$J/k_{\rm B}$\,(K) & & 123.1(1)* \\
$\delta$ &&  0.0690(3)* \\
$\Delta/k_{\rm B}$\,(K) & & 33.4(2)* \\
${\chi^2\over {\rm DOF}}$ $\big({\rm 10^{-5}{cm^3\over mol\,V}}\big)^2$
& & 2.1 \\
$\sigma_{\rm rms}$ (\%) & & 1.43 \\
\end{tabular}
\label{Table}
\end{table}
\noindent Table~\ref{Table} along with the goodnesses of fit.  The spin
susceptibilities for the three crystal directions are plotted versus
temperature in Fig.~\ref{HPVOPOXtalChiabc} as dashed curves.

The average of the three fitted $C_{\rm imp}$ values is $3.9 \times 
10^{-3}\,{\rm cm^3\,K/mol\,V}$, which is equivalent to the contribution of
1.0\,mol\,\% with respect to V of paramagnetic species with $S = 1/2$ and
$g = 2$.  This contribution is about a factor of three smaller than that
found for the powder sample in the previous section.  The average of the
three $\theta_{\rm imp}$ values is $-2.0$\,K, about the same as for the
powder sample.  The negative sign of $\theta_{\rm imp}$ may indicate AF
interactions between the defect and/or impurity magnetic moments.  The
$\theta_{\rm imp}$ can also arise from single-impurity-ion CEF effects,
and/or as a reflection of partial $T$-dependent \ paramagnetic \
saturation \ of \ the \ paramag-
\widetext
% Figure 15
\begin{figure}
\epsfxsize=5in
\centerline{\epsfbox{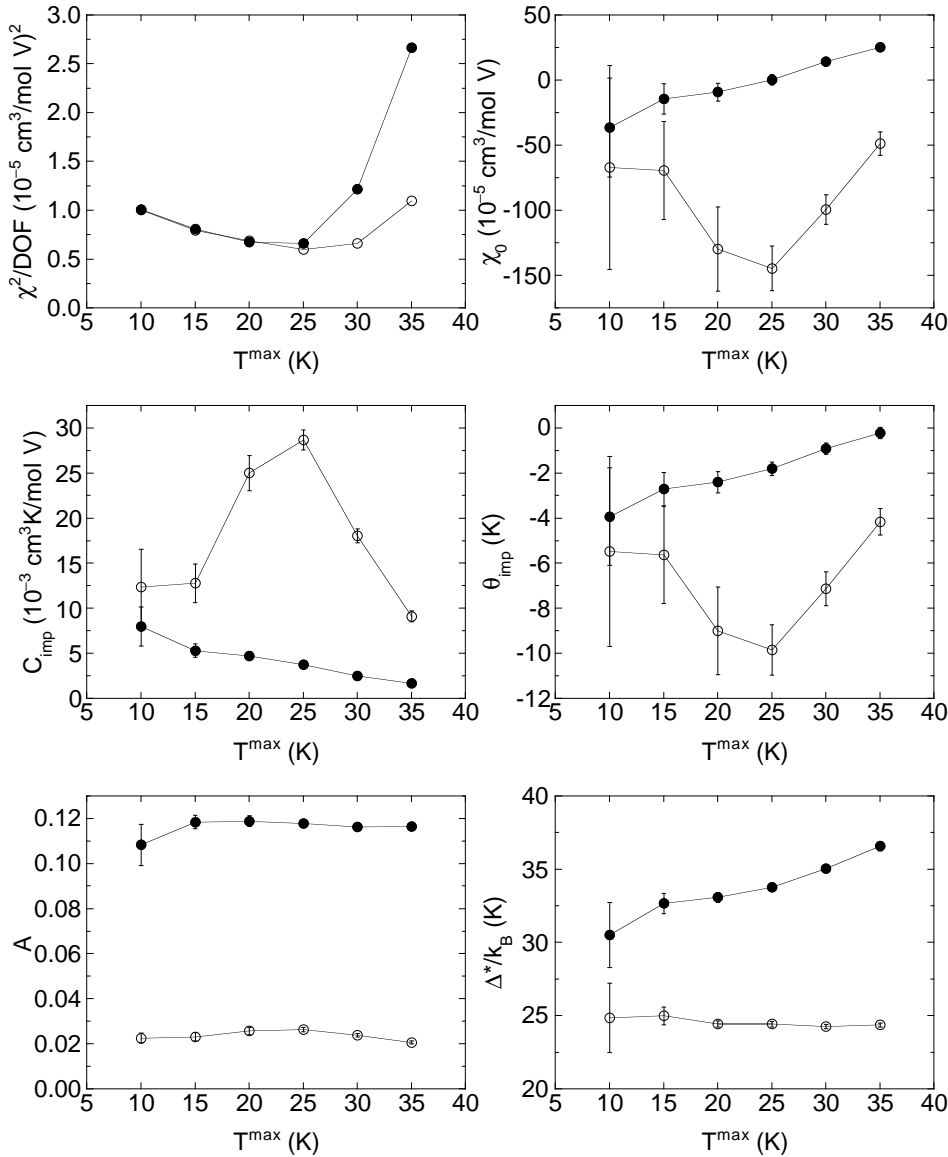}}
\vglue 0.1in
\caption{Parameters obtained from fitting the powder-average $\chi(T)$
data from Fig.~\protect\ref{HPVOPOXtalChiabc} for single crystal
HP-(VO)$_2$P$_2$O$_7$ from 2\,K up to a temperature $T^{\rm max}$ by
Eqs.~(\protect\ref{EqChiExp:all}) using the expressions for the
low-$T$ spin susceptibility $\chi^*(t)$ in Eqs.~(\protect\ref{EqTroyer:a})
($\circ$, ``Fit~1'') and~(\protect\ref{EqBul}) ($\bullet$, ``Fit~2''),
respectively.}
\label{HP-VOPOPowderAvePars}
\end{figure}
\narrowtext
\noindent  
\newpage
\noindent netic impurities at low temperatures in the fixed
field of the measurements.  The anisotropic $\chi^{\rm VV}$ values are
derived from Eqs.~(\ref{EqChiExp:b}) and~(\ref{EqChiCore}) using the
fitted anisotropic $\chi_0$ 
values in Table~\ref{Table}.  The resulting
three $\chi^{\rm VV}$ values are listed in Table~\ref{Table}.

From the above global fit to the anisotropic $\chi(T)$ data for all three
field directions, we find $J_1/k_{\rm B} = 131.6(1)$\,K and $\alpha =
0.8709(5)$.  The average of the two exchange constants $J_1$ and $J_2$
according to Eq.~(\ref{EqDimParams:d}) is $J/k_{\rm B} = 123.1(1)$\,K,
and the alternation parameter expressed in the form of $\delta$ is
obtained using our $J_1$ and $\alpha$ values and
Eq.~(\ref{EqDimParams:c}) as $\delta = 0.0690(3)$.  Using our fitted
$\alpha$ and $J_1$ parameters, Eq.~(\ref{EqDimParams2:a}) yields the spin
gap  $\Delta/k_{\rm B} = 33.2(2)$\,K\@.  
Using the derived $J$ and
$\delta$ parameters, the more accurate  Eqs.~(\ref{EqD(d):all}) \ predict
\ the \ spin \ gap \ to \ be \ $\Delta/k_{\rm B} = 33.4(2)$\,K, 

\newpage
\noindent which is the same to
within the error bars as the first estimate.  

A summary of all the fitted and derived quantities obtained in this
section from our modeling of the $\chi(T)$ measurements for single
crystal HP-${\rm (VO)_2P_2O_7}$ is given in Table~\ref{Table}.

\subsubsection{Powder average of single crystal $\chi(T)$: low-$T$ fits}
\label{SecChiPwdAve}

Additional fits to the powder-averaged single crystal $\chi(T)$
data were carried out at low temperatures to obtain an estimate of the
spin gap which is independent of the model for the gapped spin
susceptibility.  The powder average was used in order to reduce the number
of parameters needed to fit the data.  We used the general fit
expressions~(\ref{EqChiExp:all}) in which the spin susceptibility
$\chi^*(t)$ is given by the low-$T$ approximations in
Eq.~(\ref{EqTroyer:a}) (``Fit~1'') or Eq.~(\ref{EqBul}) (``Fit~2''), and
where the prefactor parameter $A$ was fitted independently of the spin
gap.  The only difference between Fits~1 and~2 is the  exponent of $1/t$
in the prefactor to the exponential in the expression for the low-$T$
$\chi^*(t)$, which is 1/2 for Fit~1 and~1 for Fit~2.

The powder-averaged single-crystal $\chi(T)$ data from 
Fig.~\ref{HPVOPOXtalChiabc} were fitted from 2\,K up to a maximum
temperature $T^{\rm max}$, and the fitted parameters obtained from
Fits~1 and~2 are plotted versus $T^{\rm max}$ in
Fig.~\ref{HP-VOPOPowderAvePars} as open and filled circles,
respectively.  Also shown in Fig.~\ref{HP-VOPOPowderAvePars} are the
statistical variances ($\chi^2$ per degree of freedom) obtained from both
types of fits.  The variances for both fits are quite similar and both
have a minimum for $T^{\rm max}\approx 25$\,K\@.  However, the impurity
Curie constant and Weiss temperature for Fit~1 in
Fig.~\ref{HP-VOPOPowderAvePars} are strongly and nonmonotonically
dependent on $T^{\rm max}$ in contrast to the corresponding dependences
for Fit~2.  In addition, the values of the fitted $\chi_0$ values for
Fit~1 in Fig.~\ref{HP-VOPOPowderAvePars} are all strongly negative.  Since
$\chi_0$ cannot be more negative than $\chi^{\rm core}$ as estimated
above in Eq.~(\ref{EqChiCore}), because the $\chi^{\rm VV}$
in Eq.~(\ref{EqChiExp:b}) is necessarily positive, the Fit~1 fits for
all the fitted $T^{\rm max}$ values are unphysical and hence the other
parameters obtained using Fit~1 are most likely also highly inaccurate.

\begin{figure}
\epsfxsize=3.4in
\centerline{\epsfbox{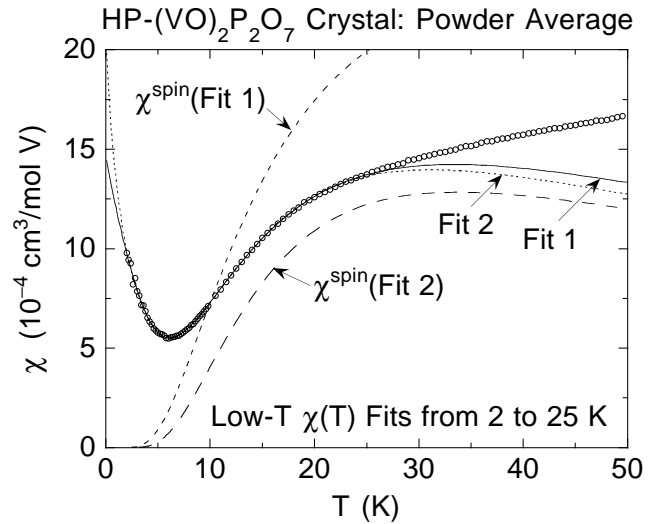}}
\vglue 0.1in
\caption{Powder-averaged magnetic susceptibility $\chi$ versus temperature
$T$ below 50\,K from Fig.~\protect\ref{HPVOPOXtalChiabc} for single
crystal HP-(VO)$_2$P$_2$O$_7$ ($\circ$).  The solid and dotted curves
are Fits~1 and~2 to the data from 2 to 25\,K using the low-$T$
approximations for the spin susceptibility in 
Eqs.~(\protect\ref{EqTroyer:a}) and~(\protect\ref{EqBul}), respectively. 
The corresponding fitted spin susceptibilities
$\chi^{\rm spin}(T)$ are shown as short- and long-dashed curves,
respectively.  Extrapolations of the curves to lower and higher
temperatures are also shown.}
\label{HP-VOPOPowderAveFit}
\end{figure}

Shown in Fig.~\ref{HP-VOPOPowderAveFit} are the two optimum fits obtained
for  $T^{\rm max} = 25$\,K for Fits~1 and~2, respectively, along with the
respective fitted spin susceptibilities $\chi^{\rm spin}(T)$.  The
$\chi^{\rm spin}(T)$ for the optimum Fit~1 is highly unlikely, as are the
fit parameters as just noted.  On the other hand, $\chi^{\rm spin}(T)$
and the fit parameters for the optimum Fit~2 are reasonable.  The values
of the fitted parameters for the \ optimum \ Fit~2 \ with
\ $T^{\rm max} = 25$\,K \ in \ Fig.~\ref{HP-VOPOPowderAvePars} \ are

%\noindent 
\[
\chi_0 = 0(4) \times 10^{-5}\,{\rm {cm^3\over mol\,V}}~,
\]
\[ C_{\rm imp} = 0.0038(4)\,{\rm {cm^3\,K\over mol\,V}}~,~~\theta =
-1.8(3)\,{\rm K},
\]
\begin{equation} 
A = 0.118(2)\,{\rm cm^3\,K\over mol\,V}~,~~{\Delta\over k_{\rm B}} =
33.8(2)\,{\rm K}~.
\label{EqChiPowdPars3}
\end{equation}
%\vglue0.33in

% Figure 16
\noindent
The first three parameters are very close to the corresponding
powder-averaged anisotropic single crystal values in Table~\ref{Table}
which were obtained in the above section assuming the $S = 1/2$ AF
alternating-chain Heisenberg model for the spin susceptibility, and
the two spin gaps are nearly identical.  The agreement between the spin
gaps found from the two independent fits supports the applicability of
this spin Hamiltonian to the spin system in \mbox{HP-${\rm
(VO)_2P_2O_7}$}.

\section{Summary and Conclusions}
\label{SecDisc}
\vglue0.07in
We have carried out detailed modeling studies of the magnetic
susceptibilities $\chi(T)$ of powder and single crystal samples of the
ambient- and high-pressure phases of (VO)$_2$P$_2$O$_7$. 
The major goal of the modeling was to determine whether the recent
proposals of a two-chain model for AP-(VO)$_2$P$_2$O$_7$
(Refs.~\onlinecite{Yamauchi1999} and~\onlinecite{Kikuchi1999}) and a
single-chain model for HP-(VO)$_2$P$_2$O$_7$ (Ref.~\onlinecite{Azuma1999})
are consistent with the respective experimental $\chi(T)$ data.  Using the
high-accuracy theoretical $\chi^*(t,\alpha)$ function for isolated
$S = 1/2$ AF alternating-exchange Heisenberg chains
in Ref.~\onlinecite{Johnston1999}, high precision tests of these models
were possible.

The $\chi(T)$ data for each phase were first analyzed using the
AF alternating-exchange chain model for {\it isolated} chains.  We found
that the proposed models are strongly supported by our high-precision
fits to the $\chi(T)$ data for each phase, from which the exchange
constants and spin gap of each type of chain in each phase were
determined.  We then considered the case of {\it coupled} chains.  The
influences of interchain couplings on the values of the intrachain
exchange constants and the spin gap of each type of chain in the two
phases were evaluated from additional fits to the $\chi(T)$ data where
the interchain coupling was treated in the molecular field
approximation.  For both phases, we find that the interchain molecular
field coupling constant is weakly ferromagnetic with a value $\lambda
\approx -0.05$, in agreement with Ref.~\onlinecite{Garrett1997a} and in
disagreement with Ref.~\onlinecite{Uhrig1998}. 
Assuming that the interchain coordination number is two and
using the values of $\lambda$ and the intrachain exchange constants,
the interchain exchange coupling constant along the $a$ axis direction
is computed to be $J_a/k_{\rm B} \approx -3.0(5)$\,K in both phases
of (VO)$_2$P$_2$O$_7$.  Thus although our modeling of $\chi(T)$, from
which the intrachain exchange constants and spin gaps were derived, did
not explictly incorporate the influence of the magnon dispersion
perpendicular to the chain direction, we believe that our mean-field
treatment effectively captures most of these effects on $\chi(T)$ since
the interchain coupling is found to be very weak compared to the
intrachain couplings.  This supposition is confirmed by our fit to the
low-$T$ powder-averaged data for a single crystal of
HP-(VO)$_2$P$_2$O$_7$ by a model-independent low-$T$ approximation for
the spin susceptibility of a 1D spin system, which yielded a spin gap of
33.8(2)\,K that is identical within the error bars to the spin gap of
33.4(2)\,K obtained from a fit to the three complete anisotropic
$\chi(T)$ data sets for the crystal using the
$\chi^*(t)$ spin susceptibility function for the alternating-exchange
chain.  The good agreement of the respective spin gaps with those
in Table~\ref{TabJAPVOPO} obtained from inelastic neutron scattering and
NMR measurements also supports the magnetic models that we have used for
the two phases.

According to the usual simple model for $d$ orbitals of transition metal
atoms in a distorted octahedral crystalline electric field, the value of
the Van Vleck susceptibility $\chi^{\rm VV}$ increases as the deviation
of the $g$ value from the free-electron value of~2 increases.  Thus from
the $g$ values determined by ESR for HP-(VO)$_2$P$_2$O$_7$ in
\mbox{Table}~\ref{TabESR}, one would predict that $\chi_{a}^{\rm VV} >
\chi_{c}^{\rm VV} \gtrsim \chi_{b}^{\rm VV}$ for this phase.  This
expectation is borne out by the values of $\chi^{\rm VV}$ in
\mbox{Table}~\ref{Table} for a high-quality single crystal of this
phase.  This agreement further supports our conclusion that the $\chi(T)$
data are consistent with the presence of a single type of $S = 1/2$ AF
alternating-exchange Heisenberg chain in this phase.  The powder average
of our $\chi^{\rm VV}$ values in Table~\ref{TabESR} is close to the value
${\rm 6\times 10^{-5} cm^3/mol\,V}$ estimated from
$\chi(T)$ and $^{31}K(T)$ NMR  measurements by \mbox{Kikuchi {\it et
al.}}\cite{Kikuchi1997} for \mbox{AP-${\rm (VO)_2P_2O_7}$}.

Additional confirmation of the two-chain model for
\mbox{AP-(VO)$_2$P$_2$O$_7$} is the agreement we find between our
predicted one-magnon dispersion relations in the chain direction for the
two chains with the results of inelastic neutron scattering measurements
at small and large wavevectors.  In the intermediate wavevector regime,
our calculated dispersion relations of the respective chains predict that
they should cross.  To our knowledge, there are no relevant inelastic
neutron scattering data yet with which to test this prediction.  With the
caveat given in the next paragraph, our final estimates of the intrachain
exchange constants and of the spin gaps of the respective
alternating-exchange chains in the two phases are given in
Table~\ref{TabJAPVOPO}, where the error bars on each quantity take our
mean-field modeling of interchain interactions into account.  

By fitting the experimental dispersion relations {\it perpendicular} to
the two chains in AP-(VO)$_2$P$_2$O$_7$ of \mbox{Garrett {\it et
al.}}\cite{Garrett1997a} by the theoretical predictions of Uhrig and
Normand\cite{Uhrig1998} which incorporate the influence of interchain
couplings, both of the couplings $J_a$ and $J_\times$ were found to be
small but nonnegligible.  In addition, the theoretical dispersion
relations\cite{Uhrig1998} show that these couplings change the spin gap
from that of an isolated chain with the same intrachain exchange
constants, whereas our modeling of the experimental
$\chi(T)$ data including the influence of the interchain coupling in a 
mean-field approximation implicitly assumed that the interchain couplings
do not change the spin gap.  Using our interchain exchange constants
obtained from fitting the experimental dispersion relations perpendicular
to the chains\cite{Garrett1997a} using their theory and using our
intrachain exchange constants obtained from modeling the experimental
$\chi(T)$ data, the spin gap of each chain was calculated using their
theory to be significantly smaller than the actual spin gap for each
chain.  Thus the intrachain exchange constants we obtain from the
mean-field treatment of the interchain coupling should be considered to be
effective values within this model.  An improved evaluation of the
exchange constants from $\chi(T)$ data will only be possible using a
theoretical expression for $\chi(T)$ which incorporates the effects of the
interchain couplings on the two spin gaps.

The spin gap of \mbox{HP-(VO)$_2$P$_2$O$_7$} obtained from analyzing
$\chi(T)$ data using the low-$T$ approximation\cite{Troyer1994} $\chi^*(t)
= (A/\sqrt{t})\exp(-\Delta^*/t)$ [Eq.~(\ref{EqTroyer:a})] is found to be
different than obtained using the above high-accuracy theoretical
$\chi^*(t,\alpha)$ function for alternating-exchange chains to analyze
the same data.  For example, from a comparison of the spin gap obtained
previously for a powder sample of \mbox{HP-${\rm (VO)_2P_2O_7}$} using
this approximation\cite{Azuma1999} with the spin gap we obtained from a
fit to the same data set using the accurate $\chi^*(t,\alpha)$ function,
we infer that the error involved in determining the spin gap using this
low-$T$ approximation is about 50\,\% in this case.  Similar
discrepancies have been found previously when analyzing $\chi(T)$ data
for 1D spin systems in a similar way.  In the compound ${\rm SrCu_2O_3}$,
for example, the spin gap of the $S = 1/2$ Cu$^{+2}$ two-leg ladders
within the  ${\rm Cu_2O_3}$ trellis layers obtained by fitting
$\chi(T)$ data up to temperatures $T \sim \Delta/k_{\rm B}$ using 
Eq.~(\ref{EqTroyer:a}) (assuming that $A$ is an independently adjustable
parameter) yielded $\Delta/k_{\rm B} = 420$\,K,\cite{Azuma1994} whereas
inelastic neutron scattering measurements on this
compound yielded $\Delta/k_{\rm B} \approx 
380$\,K.\cite{Azuma1998,Eccleston1998}  On the other hand, we found that
the low-$T$ approximation $\chi^*(t) = (A/t)\exp(-\Delta^*/t)$, in which
the power of $t$ in the prefactor to the exponential is modified, can
yield much more accurate values of the spin gap.

Our AF exchange constants in Table~\ref{TabJAPVOPO} along the
alternating-exchange V chains in the two phases of ${\rm (VO)_2P_2O_7}$
are of the same order as the nearest-neighbor exchange interactions
estimated experimentally\cite{Johnston2000} and
theoretically\cite{Korotin1999} between the V ions in the two-leg
ladder compound ${\rm MgV_2O_5}$, but are much smaller than the  value of
660--670\,K found\cite{Johnston2000,Onoda1996,Korotin1999,Miyahara1998}
for the V-V coupling across the ladder rungs in isostructural ${\rm
CaV_2O_5}$. \mbox{Korotin {\it et al.}}\ have inferred theoretically that
the large differences between the exchange constants in the latter two
compounds arise from the stronger tilting of the ${\rm VO_5}$ square
pyramids in ${\rm MgV_2O_5}$ as compared to
${\rm CaV_2O_5}$.\cite{Korotin2000}  The conventional empirical rules for
estimating the strengths of nearest-neighbor superexchange interactions
in oxides are strongly violated in ${\rm CaV_2O_5}$ and also in cuprate
spin ladder compounds, as extensively discussed in
Ref.~\onlinecite{Johnston2000}.  A similar analysis of the exchange
coupling strengths in the two phases of ${\rm (VO)_2P_2O_7}$ would be
informative and perhaps quite relevant to a more general evaluation of
this issue.

\acknowledgments
%\vglue0.33in
We thank S. E. Nagler and G. S. Uhrig for helpful discussions and
correspondence, and our recent collaborators in
Ref.~\onlinecite{Johnston1999} on work which made the present study
possible.  We are grateful to H.~Schwenk for the anisotropic
$\chi(T)$ data for a single crystal of \mbox{AP-${\rm (VO)_2P_2O_7}$} in
Fig.~4(a) of Ref.~\onlinecite{Prokofiev1998}, designated as ``crystal~2''
in the present paper, to S.~E.~Nagler for sending the
dispersion relation data for \mbox{AP-${\rm (VO)_2P_2O_7}$} in Fig.~3 of
Ref.~\onlinecite{Garrett1997a}, and to \mbox{C.~Knetter} for sending the
expansion coefficients for the dispersion relation of the frustrated 
alternating-exchange chain in Ref.~\onlinecite{Knetter2000} prior to
publication.  Ames Laboratory is operated for the U.S. Department of
Energy by Iowa State University under Contract No.\ W-7405-Eng-82.  The
work at Ames Laboratory was supported by the Director for Energy
Research, Office of Basic Energy Sciences.  This work was partly
supported by CREST (Core Research for Evolutional Science and Technology)
of Japan Science and Technology Corporation (JST) and Grant-in-Aid for
Scientific Research of the Ministry of Education, Science, Sports and
Culture, Japan.

%\newpage
%\mbox{ }
%\newpage
\vglue0.7in

\end{document}